\documentclass{aa} 
\usepackage{float}
\usepackage{tikz}
\usetikzlibrary{shapes,arrows}

\usepackage{graphicx}

\bibpunct{(}{)}{;}{a}{}{,} 
\usepackage[varg]{txfonts}

\usepackage[bookmarks=true,
  colorlinks,
  linkcolor=blue,
  urlcolor=blue,
  citecolor=blue,
  plainpages=false,
  pdfpagelabels,
  final,
  breaklinks=true]{hyperref}

\usepackage{soul}

\begin{document}

\title{Searching for a signature of turnaround in galaxy clusters with convolutional neural networks}

\author{Nikolaos Triantafyllou 
      \inst{1,2,3}\orcid{0009-0003-8609-4529}
      \and
      Giorgos Korkidis\inst{1,2}\orcid{0000-0003-3481-0880}
      \and 
      Vasiliki Pavlidou \inst{1,2}\orcid{0000-0002-0870-1368} 
      \and 
      Paolo Bonfini \inst{1}\orcid{0000-0003-1957-5878} 
      }
      
\authorrunning{N. Triantafyllou et al.}
\titlerunning{Searching for a signature of turnaround in galaxy clusters with CNNs}

\institute{Department of Physics and Institute for Computational and Theoretical Physics, University of Crete, 70013 Heraklio, Greece\\
          \email:  gkorkidis@physics.uoc.gr
          \and             
          Institute of Astrophysics, Foundation for Research and Technology – Hellas, Vassilika Vouton, 70013 Heraklio, Greece
          \and 
          Scuola Normale Superiore, Piazza dei Cavalieri 7, 56126 Pisa (PI), Italy \\ \email: nikolaos.triantafyllou@sns.it}

\date{}

\abstract 
    {Galaxy clusters are important cosmological probes that have helped to establish the $\Lambda$ cold dark matter paradigm as the standard model of cosmology. However, recent tensions between different types of high-accuracy data 
    highlight the need for novel probes of the cosmological parameters. Such a probe is the turnaround density: the mass density on the scale where galaxies around a cluster join the Hubble flow.
   To measure the turnaround density, one must locate the distance from the cluster center where turnaround occurs. Earlier work
   has shown that a turnaround radius can be readily identified in simulations by analyzing the 3D dark matter velocity field. However, measurements using realistic data face challenges due to projection effects. }
   {This study aims to assess the feasibility of measuring the turnaround radius using machine learning techniques applied to simulated idealized observations of galaxy clusters.}
   {We employed N-body simulations across various cosmologies to generate galaxy cluster projections. Utilizing convolutional neural networks, we assessed the predictability of the turnaround radius based on galaxy line-of-sight velocity, number density, and mass profiles.}
   {We find a strong correlation between the turnaround radius and the central mass of a galaxy cluster, rendering the mass distribution outside the virial radius of little relevance to the model’s predictive power. The velocity dispersion among galaxies also contributes valuable information concerning the turnaround radius. Importantly, the accuracy of a line-of-sight velocity model remains robust even when the data within the $\rm R_{200}$ of the central overdensity are absent.}
   {Single-cluster turnaround radius inference from projected observables seems to be highly challenging. Future progress is likely to require statistical approaches, especially stacking, to exploit cosmological information encoded at turnaround scales.}


\maketitle

\section{Introduction}
\label{sec:introduction}

\begin{table*}[htb!]
\centering
\caption{Simulation  box size in $ h^{-1}$ Mpc, number of particles, particle mass in $ h^{-1} \rm M_\odot$, force resolution in units of $ h^{-1}$kpc and cosmological parameters.}
\label{table 1}
\begin{tabular}{c c c c c c c c}
\hline\hline
\noalign{\smallskip}

Simulation & Box & Particles & $\rm M_p$ & $\rm \Omega_m $ & $\rm \Omega_b$ & $\rm \Omega_\Lambda$ & $  h_{100}$  \\ 
\hline
\noalign{\smallskip}
MDPL2              & 1000 & $3840^3$ & $1.51 \times 10^{9}$ & $0.307115$ & $0.048206$ & $0.692885$ & 0.6777 \\
Virgo $\rm \Lambda CDM$ & $239.5$ & $256^3$ & $6.86 \times 10^{10}$& $0.3$ &       -         & $0.7 $& 0.70\\
Virgo SCDM         & $239.5$ & $256^3$ & $22.7 \times 10^{10}$ &$1.0$ &        -        & $0.0$ &$0.50$ \\
Virgo OCDM         & $239.5$ & $256^3$ & $6.86 \times 10^{10}$ & $0.3$ &       -         &  $0.0$ & $0.70$  \\
\hline
\end{tabular}
\tablefoot{All simulations are from dark-matter--only runs.}
\end{table*}

Observations of the large-scale structure in the Universe validate the $\Lambda$ cold dark matter ($\Lambda \rm CDM$) model, with the exception of some apparent contradictions, referred to as ``tensions,'' particularly in relation to observations of the Hubble parameter and \(\rm \sigma_8\)
(see e.g., \citealp{bernal2016trouble,zhao2017dynamical,joudaki-2017,hildebrandt-2017,riess-2018,riess2019large,motloch2018tensions,miville2020planck,raveri2019concordance,adhikari2019new,handley2021curvature,di2020planck,di2021investigating,shah2021buyer, vagnozzi-2023}). 
Current observational evidence for the existence of a cosmological constant is based on the relation between the present-day values of the cosmological density parameters of matter, $\rm \Omega_m$
and dark energy, $\rm \Omega_\Lambda$, probed either by the cosmic microwave background or by observations of distant supernovae. 
Other cosmological datasets (cluster abundances, baryon acoustic oscillations) are also sensitive to $\rm \Omega_m$, but much less so to $\rm \Omega_\Lambda$. 
Recent work \citep{pavlidou-2020,pavlidou-2014,korkidis-2020, korkidis2023turnaround, tanoglidis2015testing,tanoglidis2016turnaround} has demonstrated that the turnaround density ($\rm \rho_{ta}$) can be used to impose novel constraints on the cosmological parameters, highlighting that all structures share the same $\rm \rho_{ta}$ for the same redshift and cosmology.
 
The turnaround scale, also known as the turnaround radius ($\rm R_{ta}$),  marks the boundary where matter surrounding a collapsed cosmic structure transitions, on average, into the Hubble flow \citep{korkidis-2020}. Turnaround is not a singular event in a structure's history. Instead, it is a continuous phenomenon observable around all structures at any given time. For example, around a big galaxy cluster, there is a region where material is actively accreting. However, beyond a certain distance, the expansion of the Universe sweeps away other structures. The intersection of this accretion region with the Hubble flow occurs at the scale often referred to as the zero-velocity shell, the turnaround radius. 

To accurately determine $\rm \rho_{ta}$, it is essential to know both $\rm R_{ta}$ and $\rm M_{ta}$ (the mass enclosed inside the turnaround radius). 
For very nearby objects, it may be possible to reconstruct the turnaround mass and radius from the peculiar velocity field (e.g., \citealp{hoffman2018quasi}).
However, in general, these methods do not scale with distance.  The primary observational data that are widely available, even for distant clusters, are the distribution of galaxies as seen on the plane of the sky and their line-of-sight velocities. This limitation means that the turnaround radius cannot be reliably calculated by analyzing the velocity profiles directly; at a turnaround radius away from the center of the cluster as seen on the plane of the sky, the velocity component that transitions from positive to negative at turnaround is entirely on the plane of the sky, hence not observable. For the measurement to occur along the line of sight, distances of individual galaxies need to be measurable independently of their redshifts, and these are available only for nearby clusters. 
Consequently, researchers must rely on indirect methods or theoretical models to estimate the turnaround radius, such as the use of Zel'dovich sheets, as explored in \citet{lee2015bound, lee2016turning, lee2016universality, lee2017effect, lee2018estimating}.
Here, we explore a different approach and test the feasibility of using machine learning tools directly on (simulated) galaxy position, mass, and line-of-sight velocity data to identify the turnaround radius of a galaxy cluster.

The use of machine learning methods, and in particular the subcategory of deep learning algorithms, that is, artificial neural networks (NNs), has experienced an exponential growth in astrophysical and cosmological studies over the recent years (\citealp{lahav2023deep,baron2019machine,carleo2019machine,huertas2022dawes}), with several discussions on the effectiveness of these data driven methods \citep{lin2022photometric}. 
As per the universal approximation theorem, NNs---essentially compositions of weighted sums---are universal function approximators, capable of representing any continuous function in the limit of infinite parameters. Consequently, they serve as powerful tools for exploring data, with feature extraction capabilities embedded in their internal structure.
For this reason, we adopt such an approach to investigate which types of observational datasets may encode information on the $\rm R_{ta}$ of galaxy clusters.

In particular, we wish to establish whether there exists in principle a way of measuring the turnaround radius on the plane of the sky, using information on the distribution of galaxies around galaxy clusters, their masses, and their line-of-sight velocities. 
To this end, we employed N-body dark matter simulations to create projections of dense dark matter halo regions (since baryonic effects are negligible at these scales, we use the terms ``halo'' and ``galaxy'' interchangeably) and measured their turnaround radii from their velocity profiles. For all purposes of this work, we use the term ``cluster'' to refer to these overdense regions. We then applied machine learning techniques to establish a relationship between these projections and the accurate, numerically computed turnaround radii derived from the 3D information on the dark matter distribution and velocity field available in the simulations.

The outline of this paper is as follows: In Sect. \ref{sec:data} we describe
the set of N-body simulations used in this work and the characteristics of the resulting turnaround radii. 
In Sect.  \ref{sec:methods} we
describe the methods used for the creation of the model that aims to predict the turnaround radius, and in Sect.  \ref{sec:results} we present the results of our analysis. We discuss
these findings and conclude in Sect. \ref{sec:conclusions}.

\section{Data} \label{sec:data}

\subsection{N-body simulations}\label{sec:nbody}

In this work, we used data from: (a) The MultiDark Planck 2 (MDPL2) Simulation \citep{riebe2013multidark}. The MDPL2 simulation was performed with $(1 {\rm Gpc}/h)^3$ box size, with about 57 billion particles and a cosmology consistent with \cite{2014A&A...571A..16P}. It was performed using the L-Gadget2 code. (b) The Virgo Intermediate Scale Simulations of the $\rm \Lambda CDM$, open cold dark matter (OCDM), and standard cold dark matter (SCDM) models \citep{frenk2000public, thomas1998structure, jenkins1998evolution}.
These simulations contain $256^3$ particles each in a box of $(239.5 \ {\rm Mpc}/h)^3$ and represent different cosmological models. The calculations were done using the external AP3M N-body code. For a more in-depth description of the simulated boxes and the halo sample selection, the reader is referred to \cite{korkidis2023turnaround}.
For reference, we include in Table \ref{table 1} the main parameters of the four simulated boxes. 

Similarly, we used the same cluster sample as in \cite{korkidis2023turnaround} for the purposes of generating our mock cluster catalog, corresponding to the largest $\sim 3200$ 
and $\sim 1000$
halos of the MDPL2 and Virgo simulations, respectively\footnote{with a small percentage of the MDPL2 resulting from the uniform secondary sampling of \cite{korkidis2023turnaround} of logarithmically spaced bins at lower masses.}, (not making a distinction between halos and sub-halos as it is irrelevant in observations).
For each cluster in this sample, we calculated the turnaround radius kinematically by measuring the mean velocity of dark matter particles in spherical shells as described by \cite{korkidis-2020}, around the central (largest) halo.

\begin{figure}[htb!]
    \centering
    \includegraphics[width=1.01\columnwidth]{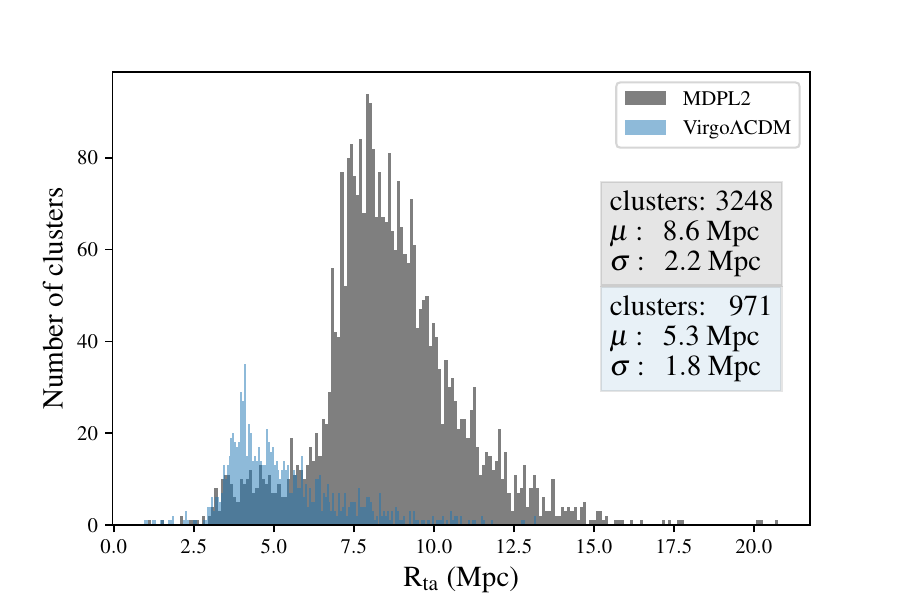}

\hfill
\caption{ 
Histogram of the turnaround radii $\rm R_{ta}$ values from MDPL2 (black) and Virgo $\rm \Lambda CDM$ (cyan). The difference between the two distributions reflects the difference between box sizes and resolutions. The mean ($\rm \mu$) and standard deviation ($\rm \sigma$) are shown on the top right, while, for the merged data $\rm (\mu, \sigma)=(7.8, 2.5)$. Additional distributions can be found in Fig. \ref{fig: Rta histogram Virgo OCDM SCDM } of the Appendix.}
\label{fig:Rta Histograms}
\end{figure}

The distribution of the turnaround radii values for the MDPL2 and Virgo $\Lambda$CDM simulations at $z=0$, along with some one-point statistics, can be seen in Fig. \ref{fig:Rta Histograms}. These shapes are expected. To first order, we expect the $\rm R_{ta}$ to depend on the central highest mass, and therefore the distribution of $\rm R_{ta}$ values should reflect the shape of the halo mass function (HMF). This implies an exponential drop at the high mass end (e.g., \citealp{Press+1974}) and a relatively hard cutoff for the low masses due to our selection of the largest halos. As this is only up to first order, deviations are observed, in part due to the secondary sampling effects in MDPL2; nevertheless, the overall expected shape is preserved. We also expect MDPL2 to more accurately capture the high-mass end of the HMF, as evidenced by the lower $\rm R_{ta}$ values from Virgo. This is a direct corollary of the fact that large-scale density Fourier modes, which, in the peak-background split formalism (\citealp{Kaiser+1984, Cole+1989, Mo+1996, Sheth+2001}) are essential for biasing regions toward collapse, cannot be represented in small volumes such as that of Virgo.

\subsection{Mock projections of galaxy clusters}

Next, we projected all galaxy-sized halos, $\rm M_{200} \geq 10^{12} M_{\odot}$ (corresponding to a stellar mass of $\rm M_\star \gtrsim 10^{10.5} M_\odot$ for $z=0$ depending on the model, see e.g., \citealp{Girelli+2020}), around each cluster, onto each of the cluster's three principal planes, retaining information on their positions, and velocities perpendicular to the planes. These velocities would be equivalent to the line-of-sight velocities as observed by an observer situated far from the cluster. Employing three projections augmented our sample size by a factor of three, a crucial benefit for the effective training of our model. The projected area for each plane was chosen to be a square region of $50 \ \rm Mpc$ on each side, centered on the respective largest halo. This dimension was selected to ensure that it would encompass the largest turnaround radius identified in our sample.

Lastly, the projection depth for the galaxies was informed by the varying uncertainties associated with different types of galaxy redshift surveys, specifically those employing either photometric or spectroscopic redshift measurements. Consequently, we created two distinct projected samples to reflect these different levels of redshift accuracy. In practical terms, this led to the imposition of a velocity cut on the velocities perpendicular to the projection plane, set at approximately $10^3 \rm km/s$ and $10^4 \rm km/s$, for spectroscopic and photometric redshift surveys, respectively. While the latter does not fully capture photometric redshift uncertainties, it is sufficient for the scope of this work; a more detailed treatment is left for future studies.

All of the aforementioned steps were carried out at $z=0$.
For the MDPL2 (Virgo $\rm \Lambda CDM$) data, we additionally considered another two redshift snapshots, namely $z=0.49$ and $z=1.03$ ($z=0.5$ and $z=1.0$). Throughout the paper, $z=0$ is implied when redshift is not explicitly stated. In the histograms of Fig. \ref{fig: Rta histogram Virgo OCDM SCDM }, we can see the distribution of the $\rm R_{ta}$ values from each of the additional MDPL2 redshift snapshots (left panel) and from each of the non-$\rm \Lambda CDM$ Virgo simulations at $z=0$ (right panel). 
The different location of peaks in different redshifts and different cosmologies is a direct indication that $\rm \rho_{ta}$ is a function of redshift and cosmological parameters.

\section{Methods}\label{sec:methods}
\subsection{Post-processing}\label{sec:prepro}

As mentioned in Sect. \ref{sec:data}, our data consists of projected regions around large dark matter halos, incorporating the details of each surrounding halo’s coordinates, virial mass, and line-of-sight velocity. This format was not suitable for our analysis (see also Sect. \ref{sec:cnns}). Hence, in order to feed the data to the algorithm, we converted them into images as 2D histograms. For the histograms,
we used appropriate weights to create three types of images: mass column density, number column density,
and mean line-of-sight velocity in each bin, in our case (also abbreviated as ``mass,'' ``vel,'' and ``num'' hereafter). We did not consider observational errors in each ``observed'' quantity. 
The reason for not adding noise is that we wanted to provide the model with the most favorable conditions possible. Should our analysis fail to identify the turnaround scale, it is then certain that it will also fail in noisy datasets. However, if we succeed in identifying the turnaround scale, clearly, further analysis is needed to quantify the effect of noise before proceeding to the analysis of real data.

For the first two cases (mass column density and number column density), the transformation did not change the nature of the data since we merely normalized by dividing by the area of each pixel (constant) to acquire the densities. For example, for the mass column density, we have that

\begin{equation}
    \{ \rm pixel_i\} = \sum_{h \; \in \; \rm halos \; in \; i} \frac{M_h}{ pixel \; area} .
\end{equation}
For the third case, by taking the mean velocity, some of the pixel values could change relative to one another in comparison with just binning the velocities, since we divided each pixel by a different number of halos, that is,
\begin{equation}
    \{ \rm pixel_i\} = \frac{1}{N_i} \sum_{h \; \in \;  halos \; in \; i} \frac{v_{los, h}}{ pixel \; area} 
    \neq
    \sum_{h \; \in \;  halos \; in \; i} \frac{v_{los, h}}{ pixel \; area},
\end{equation}

\noindent where $\rm N_i$ is the number of halos in pixel i. However, we discovered through several tests that this contributes positively to the outcome of our model. 
In addition to the original three types of images, we also considered a fourth type: the velocity dispersion in each bin, as a more promising (in terms of NN performance) and realistically observed quantity.

Different resolutions of the images were tested; we chose $25\times 25$ pixel images as optimal, both in terms of accuracy and computational efficiency, which allowed for more effective hyperparameter tuning\footnote{The effects of image resolution and smoothing on the CNN's performance can be found in Appendix \ref{sec:resolution and smoothing}.}. 
Higher resolution images had too many 0-valued pixels, which appeared to confuse the algorithm and made training with higher resolution images harder to converge, since the data are naturally sparse.
Changes in the network's structure with the use of L1 and L2 regularizations, which constrain the synaptic weights of the network, were not found to increase the high-resolution model's performance. Similarly, changes in the nature of the projected images by scaling the data (min-max scaling, standardization) or by exchanging the 0-valued pixels with the mean or the median of the training set, produced unsatisfactory results.

We also treated $\rm R_{ta}$ as a deterministic target variable, since a more detailed treatment of its uncertainty would obscure the results of our feasibility study. We saw from Fig. \ref{fig:Rta Histograms} that the turnaround radius data are not uniform but rather 
present a Gaussian distribution-like form (descending toward the right side due to the halo mass function and toward the left due to our sampling as explained in Sect. \ref{sec:data}). This could potentially result in the algorithm only
predicting values close to the peak, resulting in a problem usually referred to as mode collapse \citep{lin2022photometric}. 
We tried to transform the $\rm R_{ta}$ data so that the distribution of the values was less likely to affect the model. Namely, we tried to pass $\rm R_{ta}$ through an invertible function to transform the data into a uniform distribution. The results showed that the original data were more successful, and our final models were not affected by this issue.

To train a NN, the splitting of the data is necessary. Commonly, for each case, 80\% of the simulation
data were used for the training and validation (with a ratio of 4:1). The remaining 20\% were used for testing the trained model.

\subsection{Convolutional neural networks}\label{sec:cnns}

In the introduction of this paper, we established the potential of NNs for predicting $\rm R_{ta}$. We now focus on designing a model specifically tailored to the nature of our data. 
The observational proxies for $\rm R_{ta}$
are inherently spatial in nature, represented as 2D projections of galaxy distributions, along with their masses and their velocities.

This spatial structure necessitates a NN architecture capable of effectively capturing and analyzing such features.
In conventional NNs, the layers are fully connected, designed primarily for one-dimensional inputs. In this work,  we used convolutional neural networks (CNNs), which are best suited for single and multi-channel images (2D and 3D inputs).  CNNs use convolutional layers, which connect the neurons of one layer only to a small region of the neurons of the previous layer (e.g., pixels of the image), allowing for the recognition of low-level spatial features. Although a simple NN with fully connected layers (as a universal function approximator) would have been able to approximate any function, we expect spatial correlations in our data (since halos trace the underlying dark matter density field), which are captured more easily with CNNs.

Graph neural networks (GNNs) are a promising alternative. They operate on graphs, which consist of nodes representing entities (e.g., in our case, halos characterized by positions, masses, and velocities), edges encoding relationships between these entities, and potentially global context features (e.g., cosmological parameters or background overdensity). The graph’s connectivity, defined by the edges, plays a crucial role in the model's performance. In our case, the most natural and widely adopted approach would be to construct this graph based on the proximity of the halos. Two commonly used and effective choices are a k-nearest neighbors (k-NN) graph since it ensures connectivity between the nodes and a linking radius approach, as shown to sometimes produce better results for similar applications (e.g., \citealp{Villanueva+2022}). 
Preliminary tests using GNNs with graphs constructed with both approaches (for $\rm k\in \{5,10\}$ and a linking radius $\rm \in \{ 1,2,3\}\; Mpc$) did not show any improvement over CNNs. \footnote{We used \texttt{PyTorch Geometric} \citep{Fey+2019} on the MDPL2 dataset, using the halo mass and/or line-of-sight velocity as node features. The best GNN model consisted of two graph convolutional network (GCN) layers with 64 and 16 neurons, respectively, each followed by ReLU activation and dropout (30 \%), followed by a global mean pooling operation and a fully connected layer of eight neurons. The network was trained using the Adam optimizer \citep{kingma-2014} with a learning rate of 0.1.}

The main structure of our CNN can be seen in Table \ref{tab:cnn arch}. This structure was the best candidate in almost all of the trials performed, as it was favored by our optimization study.

\subsection{Metrics}

To evaluate the accuracy of the predicted $\rm R_{ta}$, we utilized established statistical metrics: the coefficient of determination (R$^2$ score) and Shapley values (\citealp{Shapley+1953}). Both are explained in more detail in the following paragraphs.

We used the $\rm R^2$ score to evaluate the accuracy of the model, by comparing the predicted and the true $\rm R_{ta}$, it is defined as
\begin{equation}
\rm
    R^2=1-\frac{\sum_i (R^{true}_{ta, i}- R^{pred}_{ta, i})^2}{\sum_i(R^{true}_{ta, i}-\langle R^{true}_{ta, i} \rangle)^2},
\end{equation}
where i are samples from the test set.
An R$^2$ score of 1 signifies that the model accurately predicts the target variable ($\rm R_{ta}$) based on the feature variables (projections). Conversely, an R$^2$ score of 0 indicates that the model does not predict the target variable at all; in other words, the model is no better than simply taking the mean of the target variable distribution. Notably, the R$^2$ score can become negative for arbitrarily worse models when the chosen model not only fails to predict the truth, but even predicts a number further away from the mean.

In the context of machine learning and the interpretability of its models, Shapley values provide a method to allocate the contribution of each feature (pixels in our case) to the prediction for each individual instance. 
In this work, we used the \texttt{SHAP} library (\citealp{Scott+2017}), which provides an efficient implementation of Shapley value estimation.
The main idea is to approximate the prediction of a complex black-box model $\rm f$ (in our case, a NN), with a simpler interpretable explanation model $\rm g$. 
\texttt{SHAP} uses a simplified vector of binary input variables $\rm x'\in\{0,1\}^M$, indicating the presence or absence of each of the $\rm M$ inputs. These are mapped to the real input $\rm x$ through a function $\rm x = h_x(x')$, which fills in missing features using their expected distribution. The explanation model is defined as 
\begin{equation}
\rm g(z') = \phi_0 + \sum_{i=1}^{M} \phi_i z'_i,
\end{equation}

\noindent where $\rm z'$ is a binary vector (similar to $\rm x'$), $\rm g(z')\approx f(h_x(z')) \text{ when } z' \approx x'$, and $\phi_0 = \mathbb{E}[f(x)]$ is the expectation value of the model, given a specific dataset. The only solution to the values $\phi_i$ that satisfy the desirable properties (described in \citealp{Scott+2017}) are the Shapley values:

\begin{equation}
\rm 
\phi_i(f, x) = \sum_{z' \subseteq x'} \frac{|z'|! \, (M - |z'| - 1)!}{M!} \left[ f(h_x(z'_{+i})) - f(h_x(z'_{-i})) \right],
\end{equation}

\noindent where $\rm |z'|$ is the number of non-zero elements in $\rm z'$, $\rm z'\subseteq x'$ indicates the sum over all possible $\rm z'$ that have a lower or equal number of non-zero elements to $\rm x'$, and $\rm z'_{+i}$ ($\rm z'_{-i}$) is the input vector including (excluding) feature $\rm i$.

In practice, \texttt{SHAP} quantifies how much each input feature pushes the model's prediction above or below the expected value $\rm \phi_0$. For this study, we used the \texttt{DeepExplainer}, a gradient-based variant of the \texttt{SHAP} library, to efficiently approximate Shapley values.

\section{Results}\label{sec:results}

In this section, we detail the outcomes of our analysis. Our primary focus was to assess the predictive capability of our trained CNN in determining the $\rm R_{ta}$ of simulated clusters, utilizing their various mock projection images. We specifically examined the network's effectiveness, trained on different mock data, including the mass distribution of galaxies in the cluster regions, as well as their velocity distribution, the surrounding galactic number density, and varying levels of halo projection around the clusters as viewed in the plane of the sky (initially the three types of histograms of Sect. \ref{sec:prepro} with the two velocity cuts, while considering the fourth type in Sect. \ref{subsec:velocity}). Additionally, we considered the impact of the central cluster region data on our model, and we evaluated the model's predictive accuracy when tested in a cosmological setting different from the one in which it was trained.

\subsection{General}\label{subsec:general}
    Initially, we focused on analyzing projected cluster images from the MDPL2 simulation at redshift $z=0$, applying both low and high velocity cuts to get a general idea of the potential of our methodologies. As outlined in Sect. \ref{sec:cnns}, CNNs can be utilized with both single and multi-channel inputs, enabling us to merge our three types of post-processed single-channel mock data images (mass column density, number column density, mean line-of-sight velocity) in various configurations to generate multi-channel images. For instance, combining all three data types can be thought of as an RGB image, with the R, G, and B channels representing our distinct data types.
    
    The outcomes of training the network with all possible data combinations are illustrated in the upper panel of Fig. \ref{fig:barplots}, depicted as a histogram of the $\rm R^2$ scores derived from the test set for each combination, displayed along the x axis. Training with the low-velocity-cut data yielded the highest accuracy, as evidenced by the superior $\rm R^2$ scores marked by blue bars. This result was anticipated, given the increased projection depth contamination associated with the high velocity cut.

    Models incorporating mass column density images demonstrated the highest accuracy, as indicated by the first four bar pairs, aligning with expectations that the gravitational influence of a cluster's mass correlates with the distance from the cluster's center, where the Hubble flow becomes dominant. 
    Mean velocity data did not exhibit a strong correlation with the $\rm R_{ta}$. We delve deeper into these observations in Sects. \ref{subsec:mass} and \ref{subsec:velocity}, examining the cases of mass column density and velocity in further detail. Moving forward, we exclusively consider the low-velocity-cut data, corresponding to the high-accuracy-redshift case, aiming to examine the most favorable scenario possible.

\subsection{Mass distribution}\label{subsec:mass}
    
    In the previous section, a significantly better performance of the models trained on data containing the mass was established. This 
    led us to investigate what the model’s performance would be without information on the central, more
    massive halo (i.e., without information on the mass of the central collapsed galaxy cluster). 
    
\begin{figure}[htb!]
    \centering
    \includegraphics[width=0.9\columnwidth]{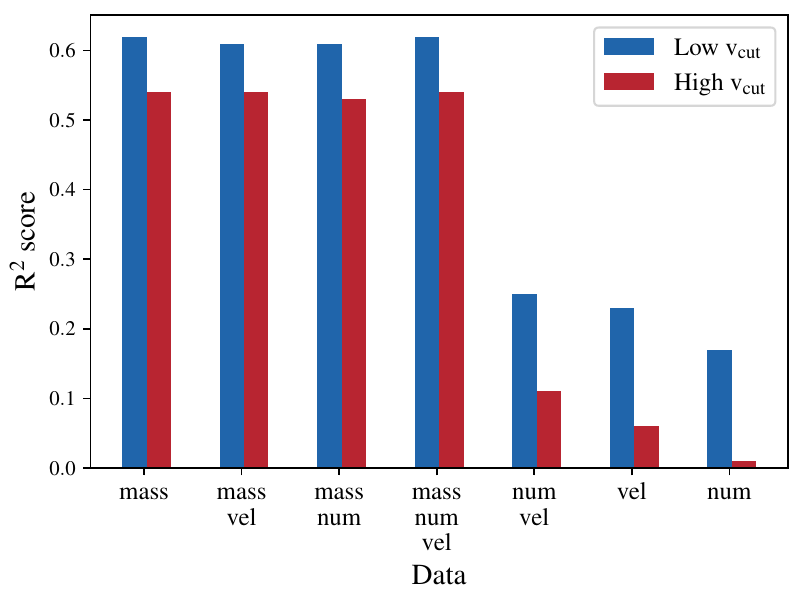}
    \includegraphics[width=0.9\columnwidth]{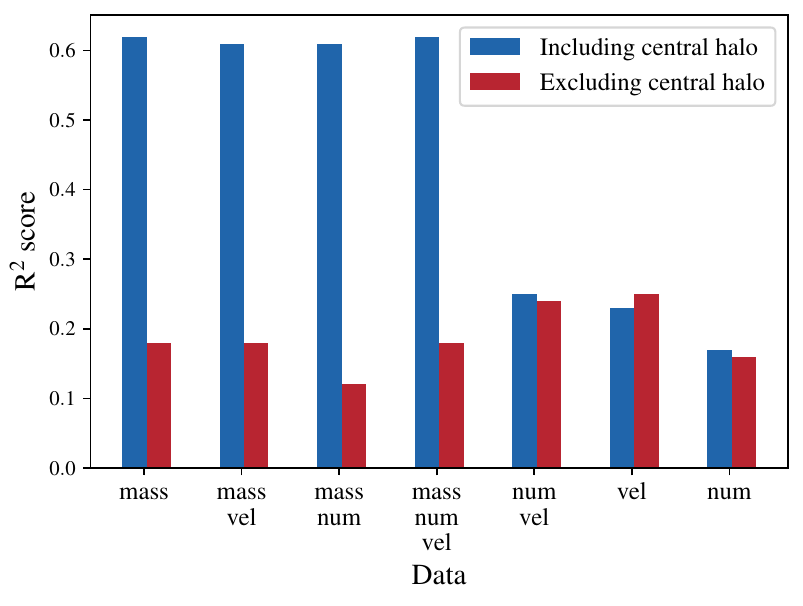}
    \hfill
    \caption{Performance comparison across different models. Upper panel: Comparison between low- and high-velocity-cut models' performance. The plot shows different $\rm R^2$ values 
    for each combination of the three features considered: mass column density, number column density, and mean line-of-sight velocity 
    (indicated as ``mass,'' ``num,'' and ``vel'' respectively). Best results are obtained for the low velocity cut (due to less contamination) and for data
    containing the mass. Lower panel: Comparison between models including or excluding the central halo. Both analyses are performed for the low-velocity-cut data.  Without the central halo, 
    performance drops significantly across models; all models without it predict the turnaround with a similar score. }
    \label{fig:barplots}
\end{figure}

    To this end, we removed the central halo completely from all of the regions and re-trained our model with the low-velocity-cut data from the MDPL2 $z=0$ output. The results can be seen in the lower panel of Fig. \ref{fig:barplots}.
    Without the central halo, the performance of the mass-containing models drops
    significantly: there is a very noticeable difference between the height of the blue and red bars in the first four models. This implies that their better performance was due to the information contained in the central halo’s mass. All of the other models predict the turnaround with about the same score.
    
\begin{figure*}[htb!]
    \centering
    \includegraphics[width=1.7\columnwidth]{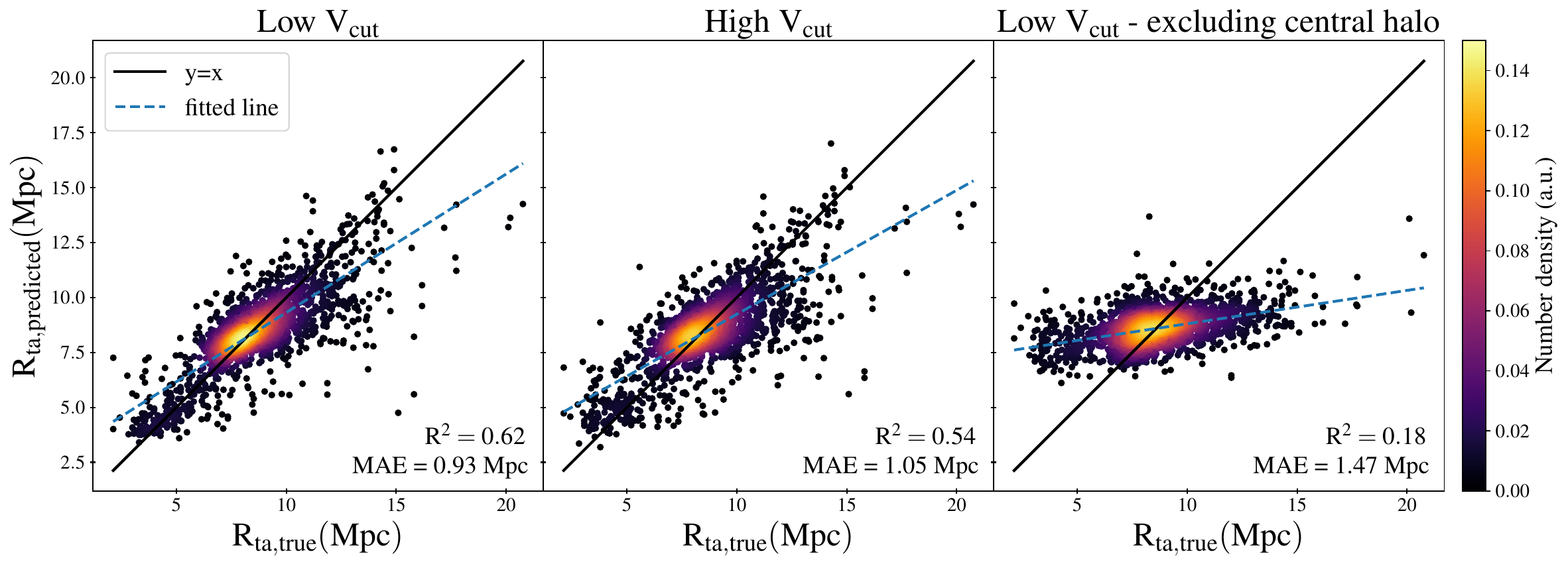}
    \hfill
    \caption{Comparison between predicted and true values of $\rm R_{ta}$ for models using the MDPL2 ``mass'' images from: the low-velocity-cut data (left panel); the high-velocity-cut data (middle panel); and the low-velocity-cut data after removing the central halo of each projection (right panel). The colors represent the number density of the plotted points, calculated using a Gaussian kernel density estimate. $\rm R^2$ scores and the mean absolute errors (MAEs) are shown at the bottom right of each plot. While the models in the left and middle panel show some correlation between true and predicted $\rm R_{ta}$ values, the right-panel model appears to mostly predict the mean value of the $\rm R_{ta}$ distribution, especially when comparing the MAEs with the standard deviation of the $\rm R_{ta}$ distribution ($\rm 2.2\; Mpc$) as shown in the upper panel of Fig. \ref{fig:Rta Histograms}.}
    \label{fig:mass true predicted}
\end{figure*}

    A more accurate representation of these results is shown in Fig. \ref{fig:mass true predicted} with
    scatter plots between the predicted values of the turnaround radii and their real values.
    If the model was perfect then all the points ($\rm R_{\rm ta,predicted}$, $\rm R_{\rm ta,true}$) would be located on the
    $\rm y=x$ (solid black) line. In the case where the central halo is missing (right panel), there seems to be little to no correlation whatsoever.
     This result underscores that what the model is extracting and learning in the first two cases (left and middle panel) is a correlation between the central mass and the turnaround radius (which has been shown to exist analytically, \citealp{Korkidis2024}) rather than some actual image feature at the location of the turnaround radius itself.

\begin{figure*}[htb!]
    \centering
    \includegraphics[width=1.5\columnwidth]{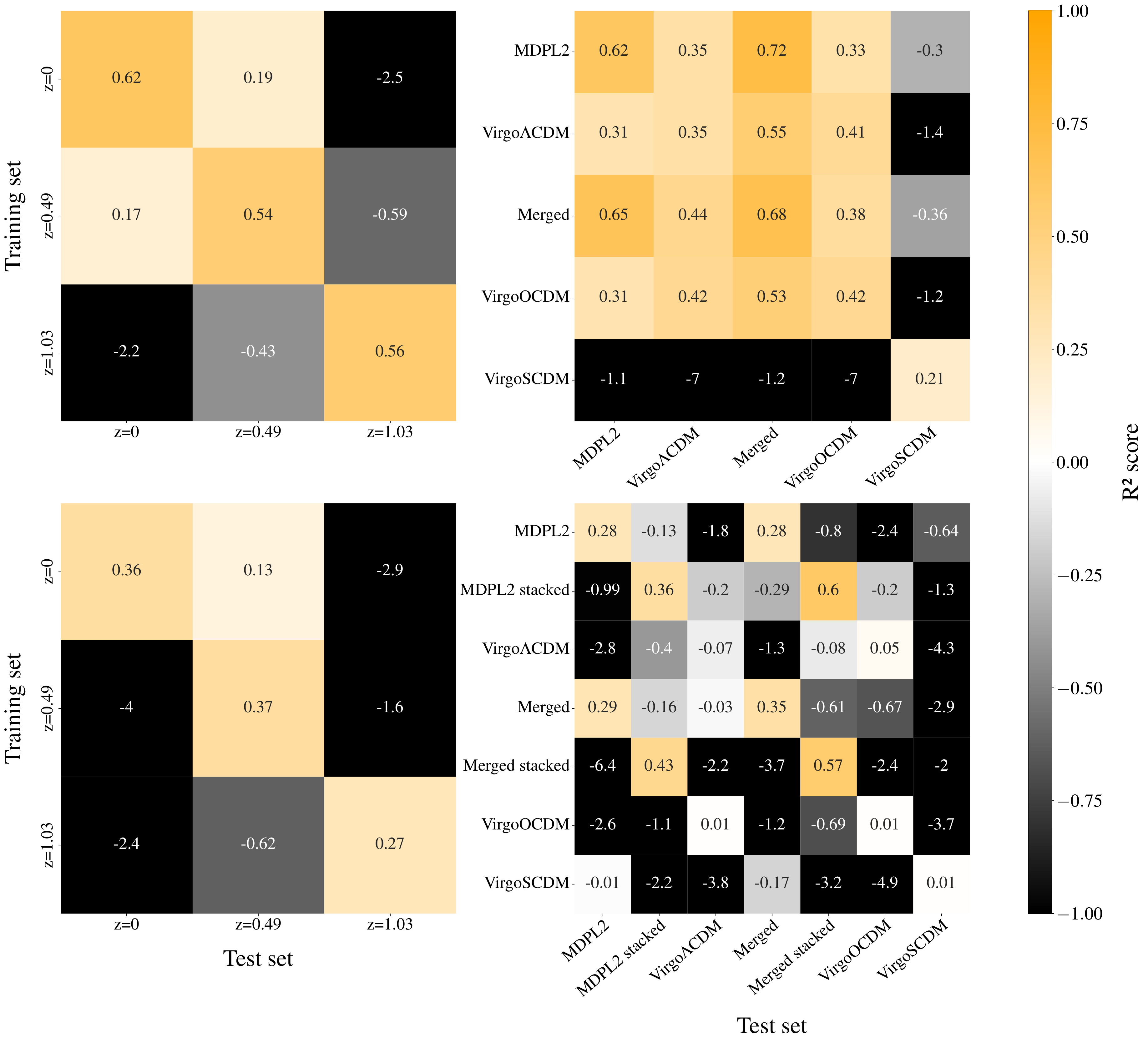}
    \hfill
    \caption{$\rm R^2$ scores of different combinations of training and testing data using the mass column density (upper panel) or velocity dispersion (lower panel) information. 80\% of the simulation
    data were used for training-validation, and 20\% was used for the testing in each case located in the diagonal. Left panel: Training and testing data from different MDPL2 redshifts (z=0, 0.49, 1.03). As expected, for each training set, the $\rm R^2$ score is the highest when evaluated on data from the same redshift. Since MDPL2 is a $\rm \Lambda CDM$ concordance cosmological simulation, the $\rm \rho_{ta}$ depends on redshift. Therefore, for different redshifts, the model cannot find an apparent correlation between $\rm M_{vir}$ and $\rm R_{ta}$ because $\rm \rho_{ta}$ varies with redshift. The velocity information plot has the same form as the mass, indicating that the model just correlates the mass with the velocity dispersion.
    Right panel: Training and testing data from different simulations and cosmological models (MDPL2, Virgo $\rm \Lambda CDM$, Virgo OCDM, Virgo SCDM, and the merged data from MDPL2 and Virgo $\rm \Lambda CDM$). Best models seem to be the ones containing the data from the MDPL2, possibly due to the fact of the larger number of training instances. Negative $\rm R^2$ values are a direct consequence of the fact that evaluating a model on another simulation can predict values even worse than the mean (as discussed in Sect. \ref{sec:methods}), since the target value distribution is an unknown.}
    \label{fig:r2 heatmap mass}
\end{figure*}

We also evaluated our model in cosmological settings different than the one our model was trained on. The $\rm R^2$ score from different redshifts from MDPL2 and different cosmologies from both MDPL2 and the Virgo simulations can be seen in the upper panel of Fig. \ref{fig:r2 heatmap mass} (left and right panel, respectively).
Best models seem to be the ones trained with the data from MDPL2, most likely due to the larger training dataset. Additionally,  a model trained at one redshift is not successful when tested at a different one. Models perform worse as the redshift of the test set moves further away from the redshift of the training set, as indicated by the decreasing $\rm R^2$ score as we move away from the diagonal in the upper left panel of Fig. \ref{fig:r2 heatmap mass}. If we accept that the model is primarily reconstructing the scaling of the central massive halo with the turnaround radius, this behavior is expected because of the redshift dependence of $\rm \rho_{ta}$, and of its scaling with the central halo mass.

    \subsection{Velocity distribution}\label{subsec:velocity}
    
    In Sect. \ref{subsec:general}, we observed that velocity data alone did not yield significant information. A model trained and tested solely on velocity data performed unsatisfactorily.  Given the known correlation between the mass of the central halo and $\rm R_{ta}$, and also between velocity dispersion and mass of the central halo, one would rather anticipate a correlation between the velocity dispersion of surrounding halos and the central halo's mass, and consequently, the turnaround radius. To address this, we explored a fourth type of data: the velocity dispersion.

    \subsubsection{Stacking}
    Our initial focus was on the MDPL2 data at $z=0$. We hypothesized that the sparsity of galaxies in computing a velocity dispersion image might obscure any relationship with the $\rm R_{ta}$. To counteract this, we employed a ``stacking'' technique for our projections, based on their respective $\rm R_{ta}$ values. This stacking involved categorizing the $\rm R_{ta}$ into bins and merging data from clusters within the same bin. We then assigned a new $\rm R_{ta}$ value for each bin, calculated as the mean of the $\rm R_{ta}$ values within it. The motivation here was that clusters normalized with respect to overdensity radii or masses are known to exhibit similar dynamical characteristics. Subsequently, as detailed in Sect. \ref{sec:prepro}, we transformed these data into images. This time, however, we focused on computing the standard deviation of the velocities within each bin.
    
    After incorporating the new data type into our training process, we reassessed the model's performance on the test set. The left panel of Fig. \ref{fig:velocity true predicted} displays the comparison between the true values and those predicted by the model, which was trained using 3000 $\rm R_{ta}$ bins (with a bin width of $\rm 0.0066\; Mpc$). This approach yielded approximately 1200 cluster images, as some bins were devoid of data. Our analysis revealed a modest improvement in correlation, as evidenced by the increase in the correlation coefficient (from $\rm R^2 = 0.24$ to $\rm R^2 = 0.36$, comparing with the one of Fig. \ref{fig:barplots}). Although this enhancement in correlation is not markedly significant, it suggests that a stacking approach, based on variables influencing cluster dynamics, holds promise for future research applications.\footnote{The results of a similar approach, but by stacking images based on their $\rm R_{200}$, can be found in Fig. \ref{fig:velocity true predicted_r200} of the appendix.}

    \subsubsection{Merging}
    From the left panel of Fig. \ref{fig:velocity true predicted}, we realized that the model was subject to the problem of mode collapse, where it just tends to predict values close to the mean of the parent distribution.
    To address the issue, we merged data from the MDPL2 and Virgo $\rm \Lambda CDM$ ($z=0$) simulations. Despite slight differences in simulation parameters, the turnaround radius should theoretically be ascertainable regardless of these minor variances. The outcomes of this merged data analysis are presented in the middle panel of Fig. \ref{fig:velocity true predicted}. The achieved $\rm R^2$ score of $0.57$, coupled with the graphical comparison of predicted versus true values, strongly suggests a significant correlation.
    
    For additional context, we analyzed various stacked images by categorizing the $\rm R_{ta}$ values into nine distinct bins. These are illustrated in Fig. \ref{fig: Stacked examples}. It is important to highlight that identifying the $\rm R_{ta}$ values based solely on these images poses a challenge. This difficulty is especially pronounced for values that fall outside the $\rm R_{ta}$ range of $4.3-10.6 \; \rm Mpc$.

    \subsubsection{Interpretability and information inside $\rm R_{200}$}
\begin{figure}[htb!]
    \centering
    \includegraphics[width=0.9\columnwidth]{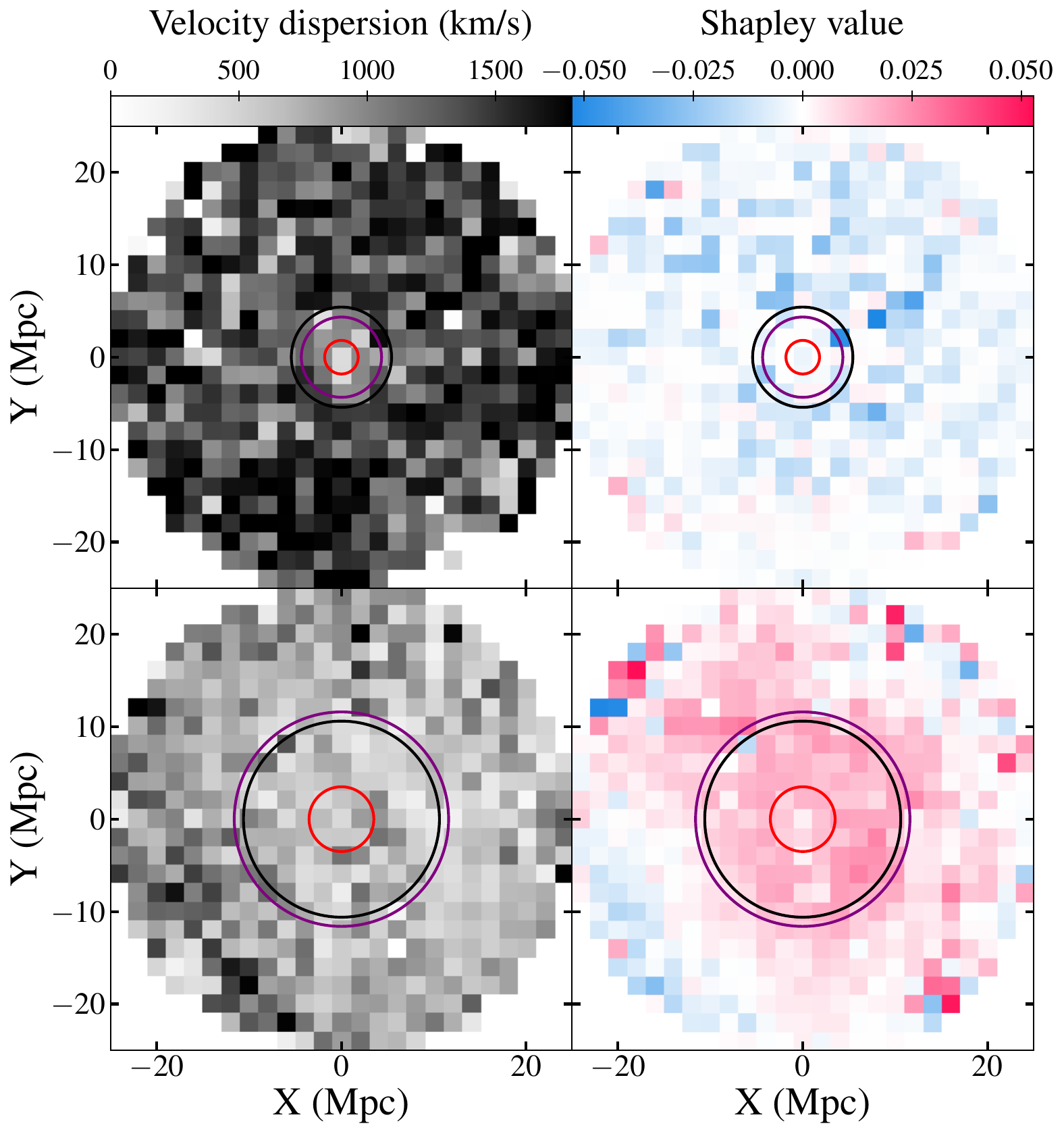}
    \hfill
    \caption{Shapley values of two random test projections (upper and lower panel) from the merged and stacked MDPL2 and Virgo $\Lambda$CDM data. Left panel: Standard deviation of the halo velocity in each pixel. Right panel: Shapley values in each pixel (pink: higher prediction; blue: lower prediction). The red circle represents the $\rm R_{200}$, while the black and purple circles represent the true and the predicted $\rm R_{ta}$, respectively. 
    Lower dispersion is always prominent in the central area due to the virialization of the matter surrounding the central perturbation. However, the extent of this lower dispersion area seems to indicate to the model where the turnaround radius should be. In particular, in the first image, the structure seems to have higher dispersion in areas far from the center, compared to the second one, which has lower values. The high dispersion seems to indicate to the model that the turnaround radius should be lower, while the lower dispersion indicates the opposite.}
    \label{fig:shapley velocities}
\end{figure}

    To better understand our findings, we analyzed the Shapley values of the pixels in a selection of test images, as depicted in Fig. \ref{fig:shapley velocities}. Our observations revealed a notable trend: a higher standard deviation in velocity in areas around the center tends to lead the algorithm to predict a lower $\rm R_{ta}$ value, and conversely, a lower velocity standard deviation suggests a higher predicted $\rm R_{ta}$. This pattern appears to support our initial hypothesis of a correlation between velocity dispersion and the mass of the central halo. However, it is crucial to note that these conclusions are drawn from just two specific examples. While Shapley values offer valuable insights, they are indicative rather than definitive. Nevertheless, a common pattern observed across most of the analyzed test images was the algorithm's apparent emphasis on the central region of the cluster in making its predictions.

\begin{figure*}[htb!]
    \centering
    \includegraphics[width = 1.7 \columnwidth]{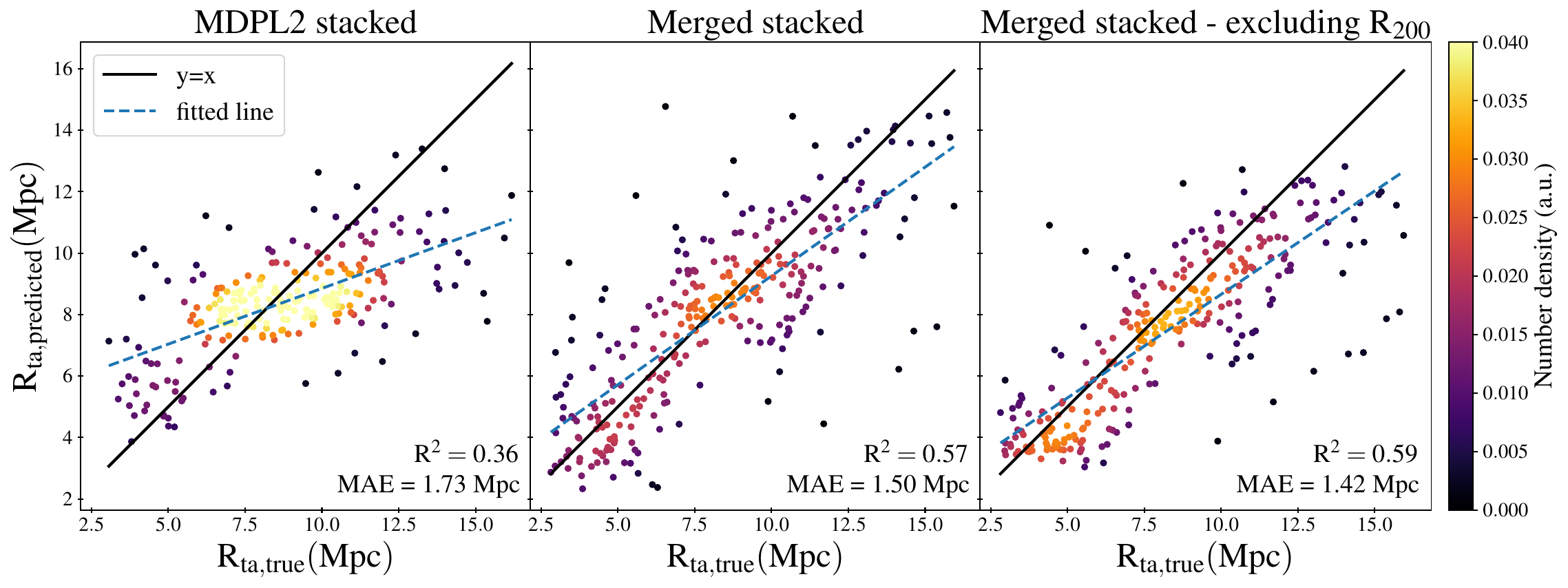}
    \hfill
    \caption{Comparison between predicted and true values of the $\rm R_{ta}$ for models using the stacked velocity-dispersion images from 3000 bins of $\rm R_{ta}$ from: the MDPL2 data (left panel); the merged MDPL2 with Virgo $\rm \Lambda CDM$ data (middle panel); the merged MDPL2 with Virgo $\rm \Lambda CDM$ data with removed information inside the $\rm R_{200}$ of the central overdensity (right panel). $\rm R^2$ scores and the mean absolute errors (MAEs) are shown at the bottom right of each plot. The colors represent the number density of the plotted points, calculated using a Gaussian kernel density estimate. It is apparent that the merging of the two datasets significantly improves the performance of the model even without the central halos' velocity information.
    }
    \label{fig:velocity true predicted}
\end{figure*}

    To further explore the CNN model's apparent focus on the central region of structures, we conducted an experiment by removing all halos within the $\rm R_{200}$ radius of each cluster's central overdensity, followed by retraining the model. The performance of this modified model is illustrated in the right panel of Fig. \ref{fig:velocity true predicted}, where it achieved an $\rm R^2$ score of $0.59$, comparable to previous results. This outcome suggests that the information within the $\rm R_{200}$ of the central overdensity is not critically important for the model's predictions when analyzing velocity dispersions.

    In a methodology parallel to that described in Sect. \ref{subsec:mass}, we conducted several training iterations with varying combinations of training and test sets across different cosmological scenarios. The $\rm R^2$ scores obtained from these experiments, involving different redshifts from MDPL2 and diverse cosmologies from both MDPL2 and the Virgo Simulations, are presented in the lower panel of Fig. \ref{fig:r2 heatmap mass} (left and right panels, respectively). Models trained with MDPL2 data consistently showed better performance, likely due to the more extensive training dataset. The $\rm R^2$ scores across different redshifts mirrored the patterns observed in the upper panel of Fig. \ref{fig:r2 heatmap mass}, which were based on the CNN trained with mass data: models trained in one specific cosmology did not predict accurately when tested in another. These results collectively suggest that the velocity dispersion information used by the CNN to predict $\rm R_{ta}$ correlates with the mass of the central overdensity. Consequently, similar to the limitations observed with mass data, the model struggles to accurately predict $\rm R_{ta}$ across different redshifts and cosmologies.

\section{Conclusions}\label{sec:conclusions}

The turnaround density of galaxy clusters has been established as a novel cosmological probe and a means to further test the current cosmological model \citep{pavlidou-2014, pavlidou-2020, korkidis-2020, korkidis2023turnaround}. In this work, we explored a method to estimate the turnaround radius of galaxy clusters based on their projected distributions on the plane of the sky. Using cosmological N-body simulations, we generated idealized observational data—incorporating mass and velocity information without uncertainties—within two ``slices'' of varying thicknesses, approximating observational uncertainties of spectroscopic and photometric surveys. The ``true'' turnaround radii were derived from the velocity profiles surrounding the central overdensities of galaxy clusters in the simulations. Then, a CNN was trained to predict the turnaround radius, utilizing various combinations of training and test data from different simulations, redshifts, and cosmologies.

When the model was trained on individual clusters, we find that the mass is the most important information we have for the prediction of the turnaround radius. This
at first may appear counterintuitive, as the turnaround is defined and was computed by the
velocity profiles of the dark matter particles around the central halo. However, it is the mass of the cluster that contributes to the gravitational dynamics defining the
turnaround radius. Furthermore, we find that the mass of the central, more massive halo was almost entirely contributing to this prediction. In the context of the findings of \cite{Korkidis2024}, this is not surprising: $\rm R_{ta}$ is an overdensity radius, so its value is entirely determined at a given redshift and for a given cosmology by the enclosed mass; and that enclosed mass has been shown analytically to scale with the central, collapsed (virial) mass of the cluster.

The velocity dispersion of the galaxies around the cluster can also be used as an information carrier, since it correlates with the enclosed mass. However, the small number of halos that were large enough to be observed within some ``neighboring'' region around a single cluster was not adequate 
for an accurate calculation of the velocity dispersion. We counteracted this effect by using stacked images combining several projections. The velocity dispersion of the halos around a cluster holds important promise for analyses of this type, and it is worth exploring further in future work, giving a clear direction to the research community for predicting $\rm R_{ta}$.

As expected, contamination from distant halos that results from the use of thicker slices affects the results. 
Thinner slices, corresponding to spectroscopic observations, would generally result in a more accurate prediction of the turnaround radius.

All of the models, when tested on data that were acquired from different redshifts and different cosmologies, were not successful in their predictions. For the models trained with the mass information, this is expected since the turnaround radius scales with the mass for a specific redshift and cosmology. For the velocity trained models the same pattern appeared, and our attempts for a straightforward rescaling that would allow the model to generalize were not successful, although they did improve performance somewhat (see appendix \ref{subsec:redshift_normalization}), so this approach with different rescaling techniques might in fact hold the most promise for a machine-learning approach to identify the turnaround scale in a cosmology-independent way. 

Nonetheless, out-of-distribution predictions are a well-known challenge in simulations, particularly in machine learning based approaches. In practice, all models and simulations contain inaccuracies, yet it is assumed they approximate reality, up to some extent. There are works trying to address this issue, for example, in \cite{Yongseok-2025} they examine the possibility of generalizing, using different cosmological simulators with different subgrid physics, for the specific task of inferring $\Omega_m$ and $\sigma_8$ from HI density slices, using a de-classifier. This is a complicated task that requires a careful examination of the latent spaces involved, using dimensionality reduction techniques, and with many caveats that are beyond the scope of this work.

While our feasibility study focused on simulations of cluster environments extended to radii of up to $\sim$20 Mpc, we acknowledge that observational data are typically more focused on a few Mpc region around clusters. To bridge the gap between simulations and observations, future efforts could focus on re-training the models using mock observations with realistic depth, noise, and sky coverage, but focusing on the central mass and the stacked velocity dispersion. This would make the networks more robust to the inherent observational effects. Moreover, extending the analysis to higher redshifts or incorporating alternative tracers may reveal additional structure; such tracers may include the line intensity mapping of interstellar medium lines (e.g., with the Spectro-Photometer for the History of the Universe, Epoch of Reionization, and Ices Explorer for H$\alpha$, the CO Mapping Array Project and the Experiment for Cryogenic Large-Aperture Intensity Mapping for CO and CII) or the 21cm line from the intergalactic medium (e.g., with the Square Kilometre Array). These observables can trace fainter galaxies and can encode information from the large-scale density field of the surrounding environment, offering complementary constraints.

Future investigations could benefit from larger and higher-resolution datasets for training, with or without stacking, addressing or ignoring the problem of model misspecification. 
A proper Bayesian treatment or bias-reduction techniques such as the one described in \cite{lin2022photometric} could become particularly useful. 
Furthermore, more complex models---for example, with the use of inception layers in the architecture of the neural network---possibly combined with another kernel choice in smoothing the density field or further exploration of GNNs, might be able to reveal additional information encoded in the mass and velocity information of galaxy clusters.

\begin{acknowledgements}
The CosmoSim database used in this paper is a service by the Leibniz-Institute for Astrophysics Potsdam (AIP).
The MultiDark database was developed in cooperation with the Spanish MultiDark Consolider Project CSD2009-00064.
The authors gratefully acknowledge the Gauss Centre for Supercomputing e.V. (\url{www.gauss-centre.eu}) and the Partnership for Advanced Supercomputing in Europe (PRACE, \url{www.prace-ri.eu}) for funding the MultiDark simulation project by providing computing time on the GCS Supercomputer SuperMUC at Leibniz Supercomputing Centre (LRZ, \url{www.lrz.de}).
The Bolshoi simulations have been performed within the Bolshoi project of the University of California High-Performance AstroComputing Centre (UC-HiPACC) and were run at the NASA Ames Research Centre.
The rest of the simulations in this paper were carried out by the Virgo Supercomputing Consortium using computers based at Computing Centre of the Max-Planck Society in Garching and at the Edinburgh Parallel Computing Centre. The data are publicly available at \url{www.mpa-garching.mpg.de/galform/virgo/int_sims}
\end{acknowledgements}

\bibliographystyle{aa}
\bibliography{bib}

\onecolumn
\begin{appendix} 

\section{Additional data distributions}
\begin{figure*}[htb!]
    \centering  
    \includegraphics[width=0.45\columnwidth]{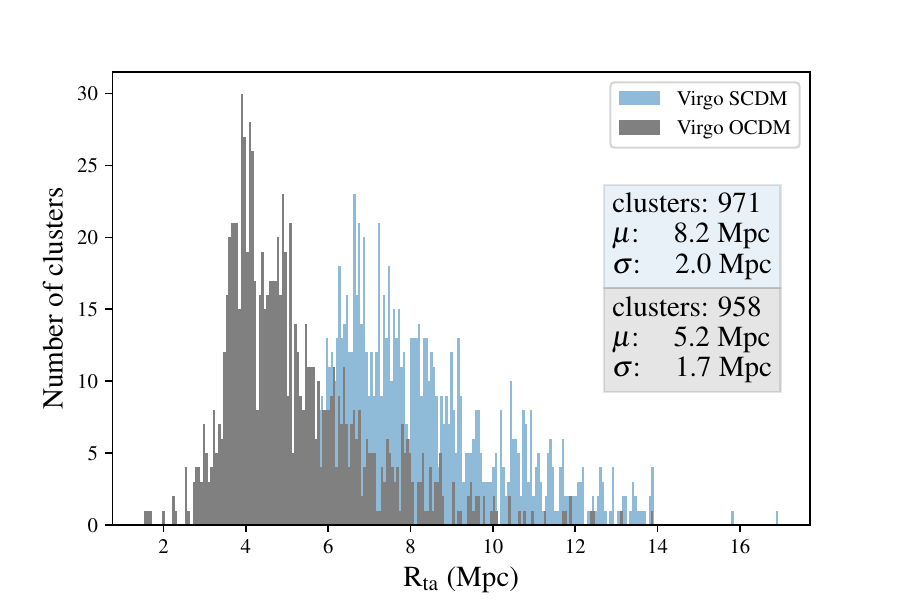}
    \includegraphics[width=0.45\columnwidth]{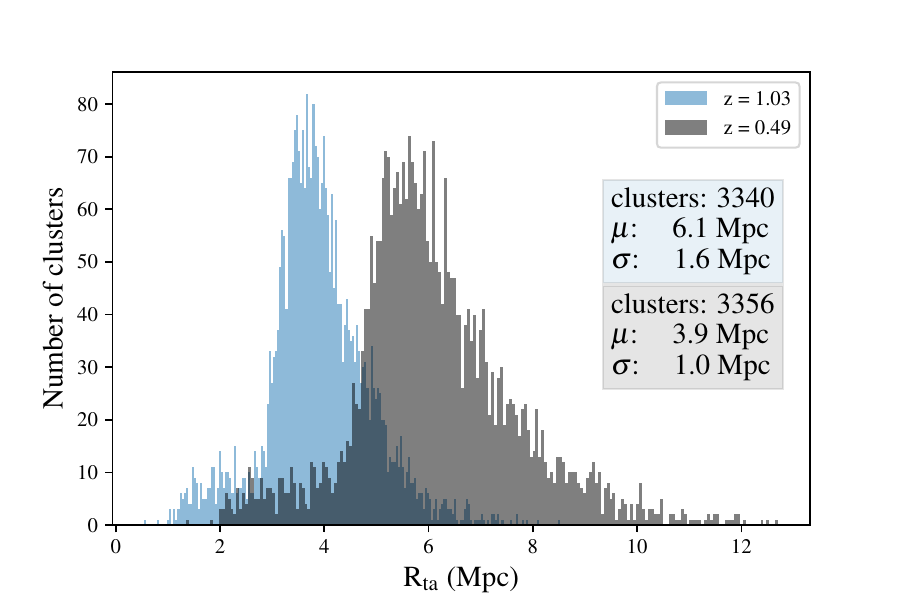}   
    \hfill
    \caption{Histogram of $\rm R_{ta}$ values from MDPL2 form $z=0.49$ and $z=1.03$ (left panel) and Virgo from OCDM and $\rm \Lambda CDM$ cosmology (right panel), similarly to Fig.\ref{fig:Rta Histograms}. The distribution moves as we change the redshift or $\rm \Omega_m$ and $\rm \Omega_\Lambda$, which is expected due to the dependence of $\rm \rho_{ta}$ on the cosmological setting discussed throughout the paper and proved in \cite{pavlidou-2020}.}
    \label{fig: Rta histogram Virgo OCDM SCDM }
\end{figure*}

 In Fig. \ref{fig: Rta histogram Virgo OCDM SCDM } we present the distributions of the $\rm R_{ta}$ values from each of the additional MDPL2 redshift snapshots (left panel) and from each of the non-$\rm \Lambda CDM$ Virgo simulations (right panel), created with the same methods mentioned in Sect. \ref{sec:nbody}.
As expected, the distributions, even though similar in terms of general form, span over different $\rm R_{ta}$ values.
In particular, the location of peaks changes between different redshifts and different cosmologies, as a direct result of $\rm \rho_{ta}$ being sensitive to redshift and cosmological parameters. If a model that successfully predicts the $\rm R_{ta}$ on all of the different regimes was created, it would mean the existence of an independent way of constraining $\rm \rho_{ta}$ and therefore $\rm \Omega_m$ and $\rm \Omega_\Lambda$.

\section{Architecture of the neural network}\label{sec:nn arch}

The NN architecture used can be seen in Table. \ref{tab:cnn arch} (implemented using \texttt{Tensorflow}, \citealp{tensorflow2015-whitepaper}).
During training, the mean absolute error was used to evaluate the model's performance on the validation set. 
In addition to this basic architecture, the hyperparameters were tuned in each case to obtain the best result. 
In most cases, the following additional elements were used: Activation function: ReLU, Padding: same, Kernel initializer: He initialization, Learning rate: $10^{-3}$, optimizer: Adam \citep{kingma-2014}. In some of the cases, L1 and L2 regularizations were also implemented.

\begin{table*}[htb!]
    \setlength{\tabcolsep}{14pt}
    \centering
    \caption{Characteristics of each layer of the CNN architecture.}
    \label{tab:cnn arch}
    \begin{tabular}{c c c c c}
    \hline\hline
    \noalign{\smallskip}
    
    Layer & Inputs & Kernel size & $h\times w$ & \#feature maps \\ 
    \hline
    \noalign{\smallskip}
    Convolution (C1)           & Input image & $5\times 5$ & $25\times 25$ & $32$ \\
    Activation  (A1)         & C1 & - & $25\times 25$ & $32$ \\
    Batch Normalization (BN1)           & A1 & - & $25\times 25$ & $32$ \\
    Pooling (P1)           & BN1 & $2\times 2$ (stride 1 pix) & $12\times 12$ & $32$ \\
    Dropout (D1)           & P1 & - & $12\times 12$ & $32$ \\
    Convolution (C2)           & D1 & $5\times 5$ & $25\times 25$ & $64$ \\
    Activation  (A2)         & C2 & - & $25\times 25$ & $64$ \\
    Batch Normalization (BN2)           & A2 & - & $25\times 25$ & $64$ \\
    Pooling (P2)           & BN2 & $2\times 2$ (stride 1 pix) & $12\times 12$ & $64$ \\
    Dropout (D2)           & P2 & - & $12\times 12$ & $64$ \\
    Flatten (F1)           & D2 & - & 2304 & - \\
    Fully Connected (FC1)           & F1 & - & 32 & - \\
    Activation  (A3)        & FC1 & - & 32 & - \\
    Batch Normalization (BN3)           & A3 & - & 32 & - \\
    Dropout (D3)           & BN3 & - & 32 & - \\
    Fully Connected (FC2)           & D3 & - & 16 & - \\
    Activation  (A4)        & FC2 & - & 16 & - \\
    Dropout (D4)           & A4 & - & 16 & - \\
    Fully Connected (FC3)           & D4 & - & 1 & - \\
    \hline
    \noalign{\smallskip}
    \end{tabular}
    \tablefoot{
    Columns are: name of the layer, input layer, size of the convolution kernel (in pixels), size (height $\times$ width in pixels), and number of the resulting feature maps.}
\end{table*}

\section{Sensitivity to image resolution and smoothing}\label{sec:resolution and smoothing}
\begin{figure}[hbt!]
    \centering
    \includegraphics[trim=0cm 0 0 0,width=0.48\linewidth]{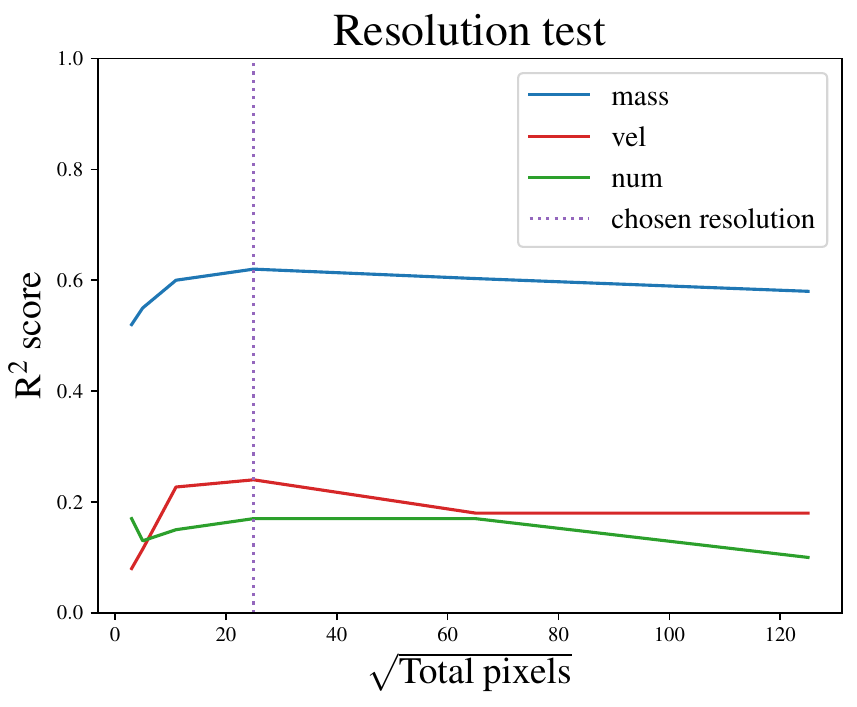}
    \includegraphics[width=0.48\linewidth]{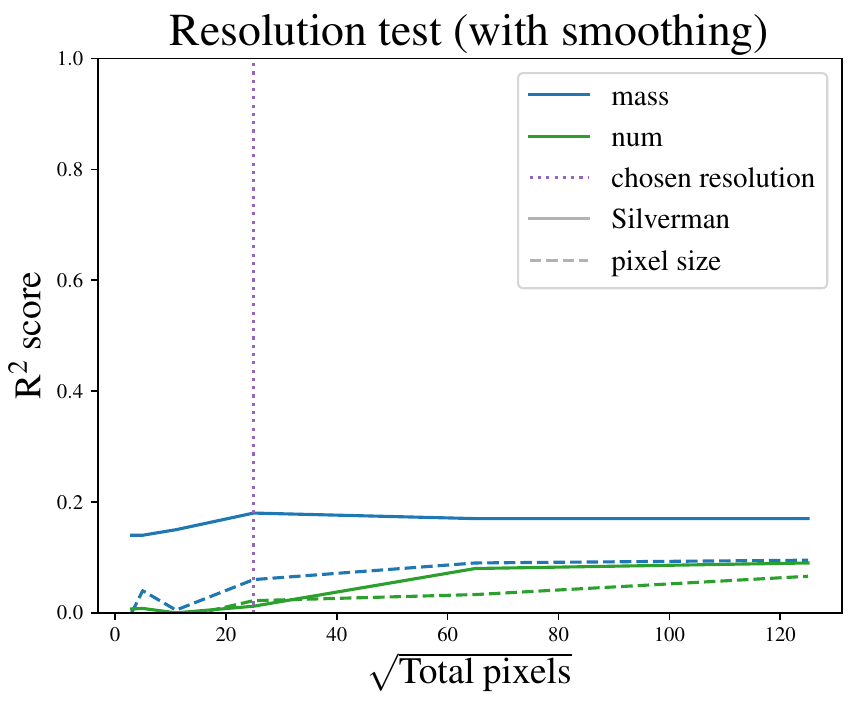}
    \includegraphics[trim=-4.8cm 0cm 6.2cm 0, width=0.40\linewidth]{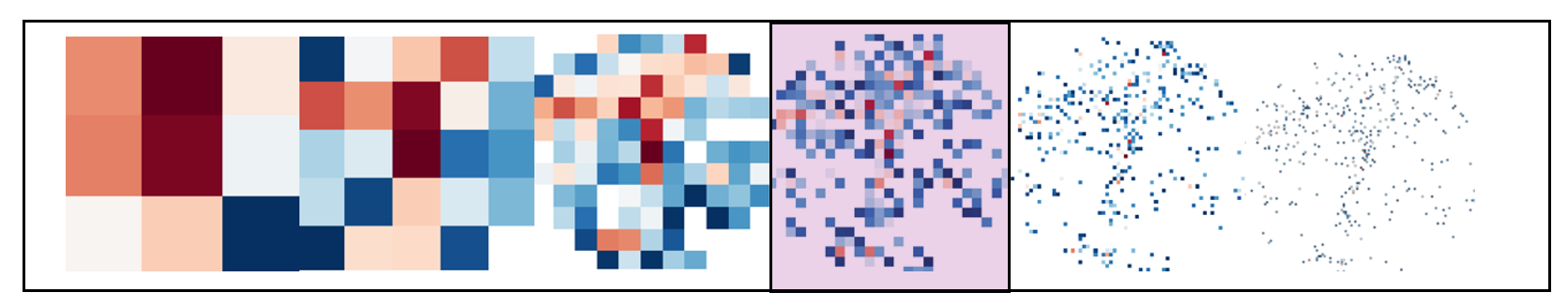}
    \hfill
    \includegraphics[trim=3cm 0 -1.8cm 0, width=0.40\linewidth]{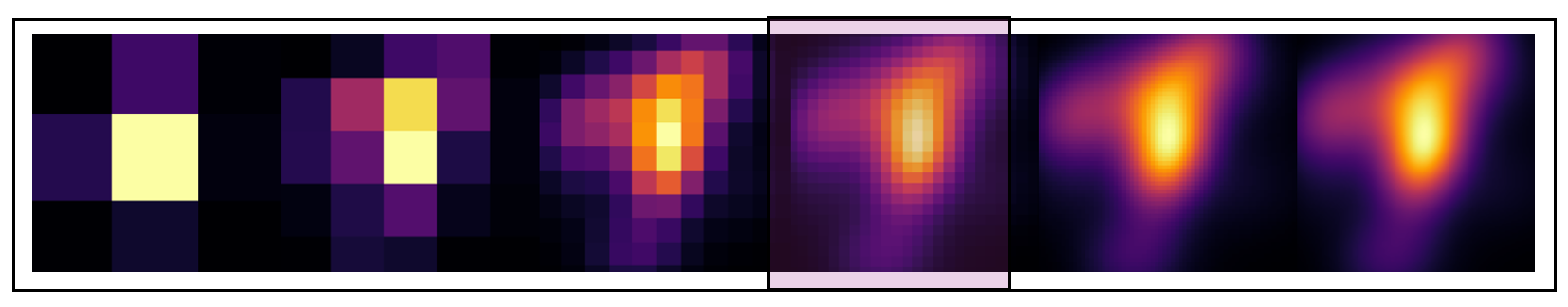}
    \caption{Resolution sensitivity test results for the base CNN model and its variations. Left panel: $\rm R^2$ scores of the best-performing model for each tested resolution ($\sqrt{\text{Total pixels}} \in {3,5,11,25,65,125}$), with example “mass” images shown below. Right panel: Same test applied to images smoothed with a Gaussian kernel density estimate, using either Silverman's rule of thumb (\citealp{silverman+1986}) or a bandwidth equal to the pixel size. Example ``mass'' images for the former are shown below. The chosen resolution (dotted purple line and shaded purple region) provides an optimal balance between predictive accuracy and numerical stability.
    }
    \label{fig:resolution-smoothing}
\end{figure}

Using the MDPL2 data (``mass,'' ``vel,'' and ``num''), we investigated the impact of image resolution and smoothing on the CNN's performance. 
To assess the effect of resolution, we tested $48$ variations of the main NN architecture of Appendix \ref{sec:nn arch} with  $ \rm {number\; of \; feature \; maps} \in \{16, 32, 64\}$, $(h,w) \in \{(2,2), (3,3), (5,5), (10,10)\}$ for both of the convolutional layers in the network and
$\rm pooling \;kernel \;sizes \in \{(1,1), (2,2), (3,3), (5,5)\}$. For each resolution, the best $\rm R^2$ score achieved across these variations is shown in the left panel of Fig. \ref{fig:resolution-smoothing}. 
The performance remains relatively stable across resolutions, and a resolution of $25 \times 25$ pixels was selected as an optimal, balancing numerical stability and information content. To evaluate the impact of smoothing, we repeated the analysis using images smoothed with a Gaussian kernel density estimate, with bandwidths set either by Silverman’s rule of thumb (\citealp{silverman+1986}) or by the pixel size (right panel of Fig. \ref{fig:resolution-smoothing}). Smoothing did not improve the performance of the CNN. This was expected since (a) in an ideal scenario, the CNN would be able to learn its own smoothing, if beneficial, and (b) the intrinsic Poissonian nature of the data is lost, reducing information about the number of collapsed structures.

\newpage 

\section{Velocity dispersion}
\begin{figure*}[htb!]
    \centering
    \includegraphics[width=0.30\textwidth]{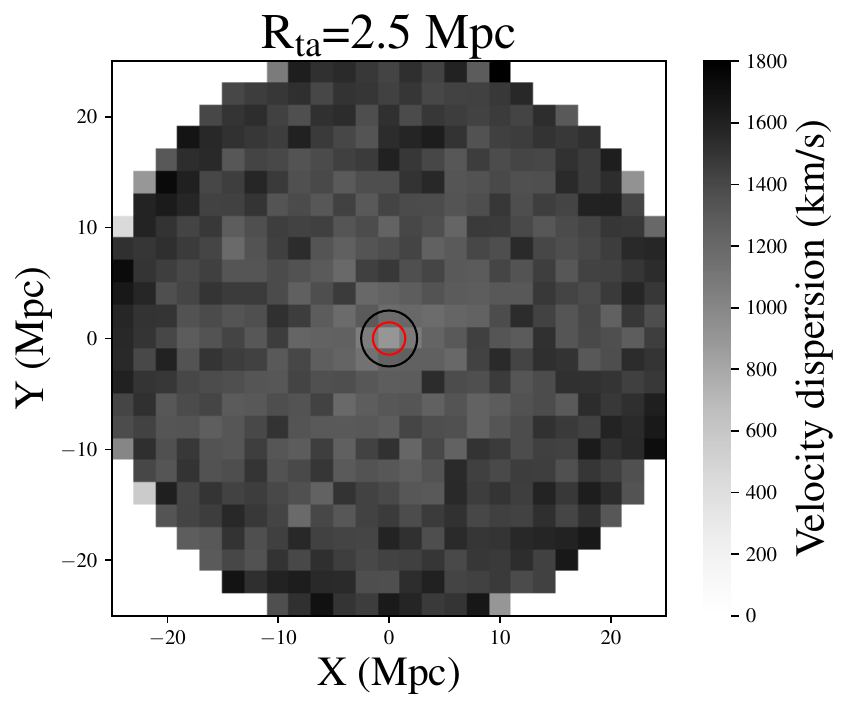}\hfill
    \includegraphics[width=0.30\textwidth]{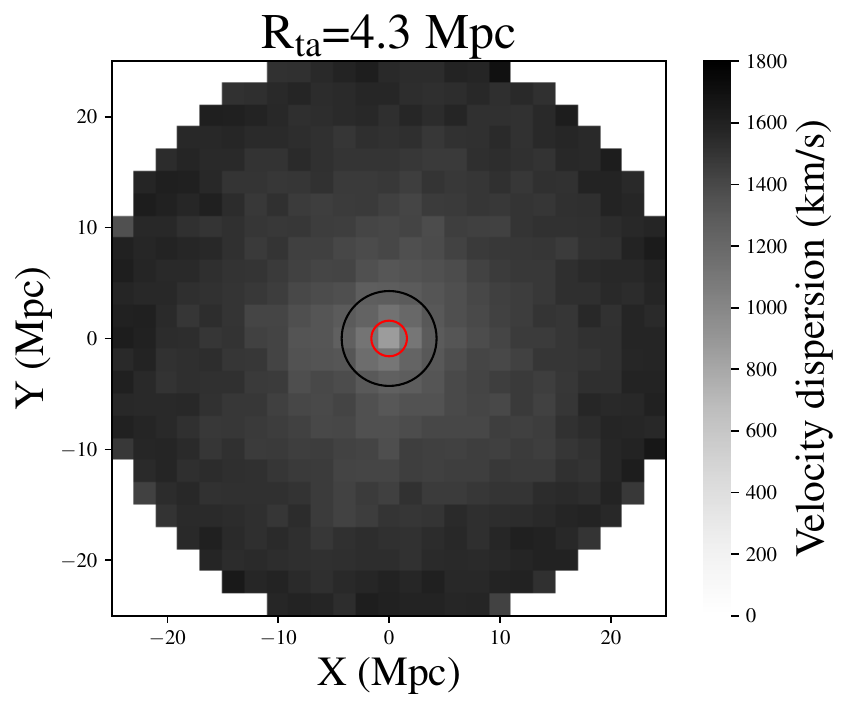}\hfill
    \includegraphics[width=0.30\textwidth]{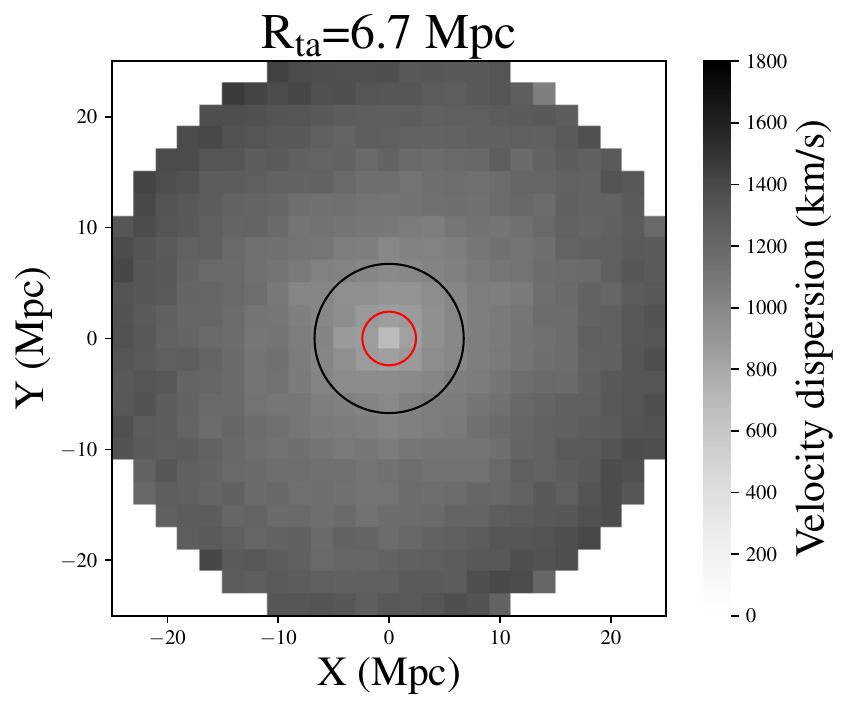}
    
    \includegraphics[width=0.30\textwidth]{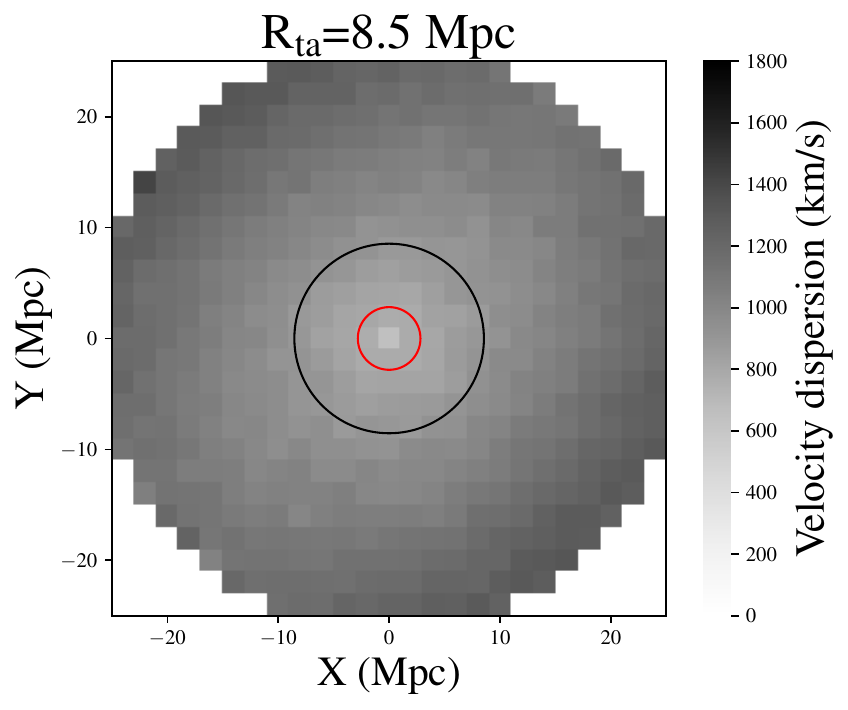}\hfill
    \includegraphics[width=0.30\textwidth]{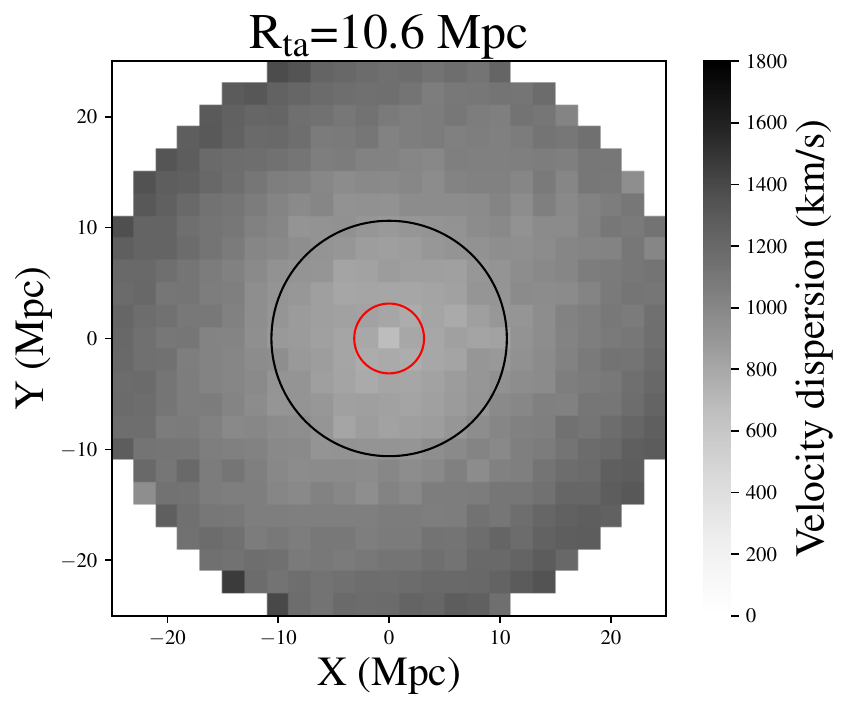}\hfill
    \includegraphics[width=0.30\textwidth]{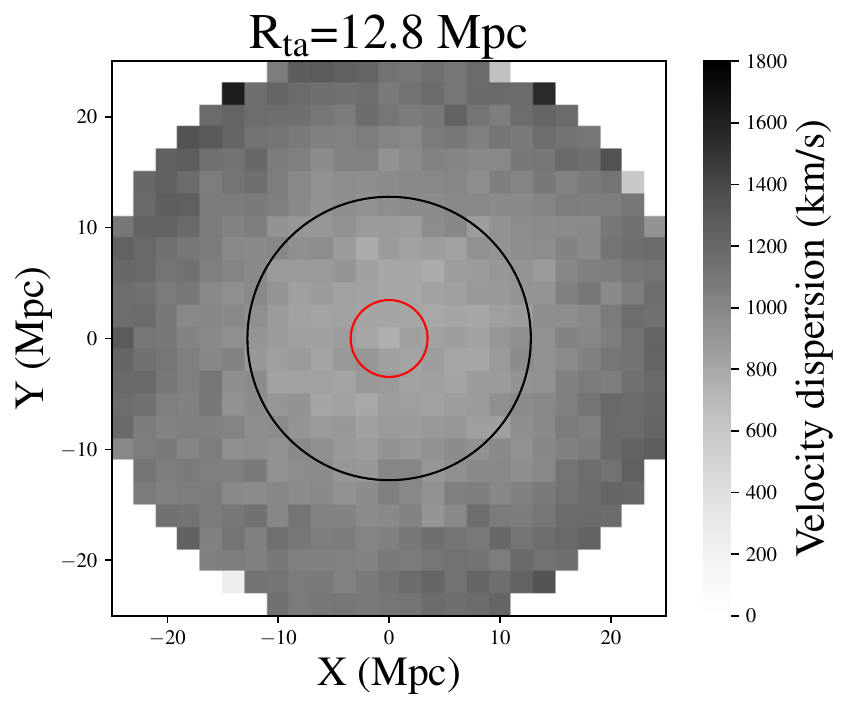}
    
    \includegraphics[width=0.30\textwidth]{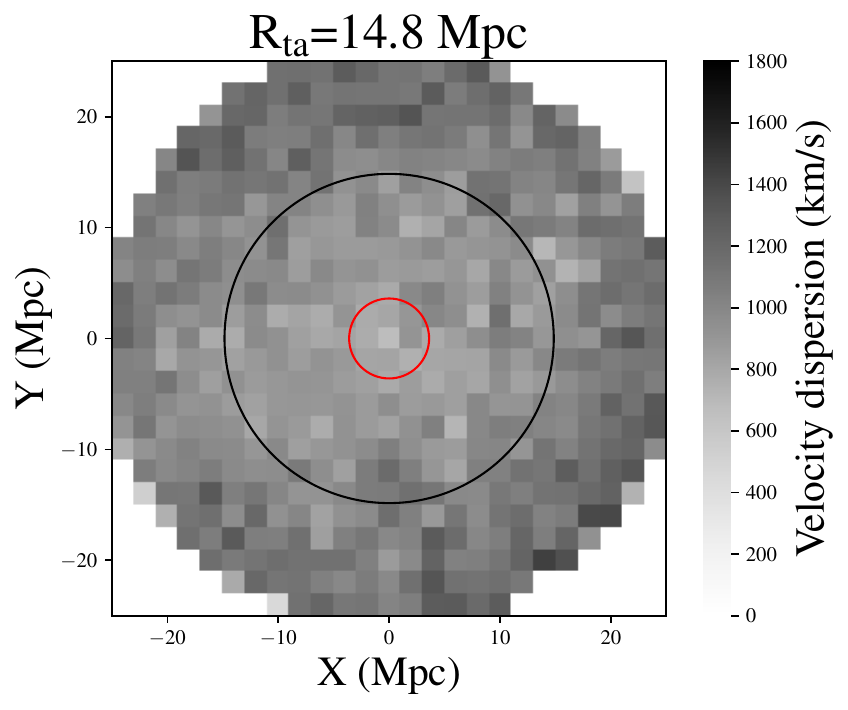}\hfill
    \includegraphics[width=0.30\textwidth]{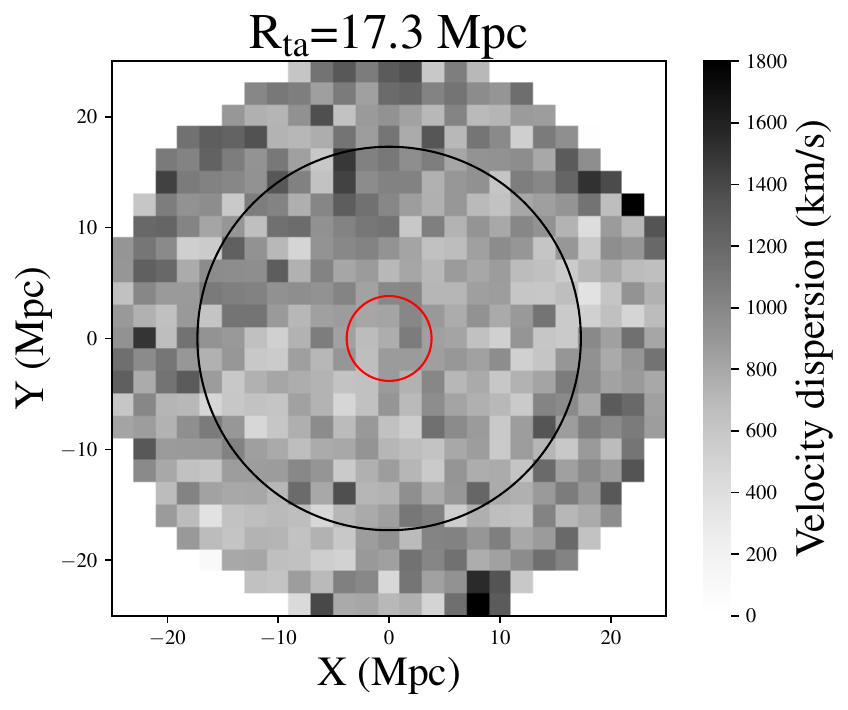}\hfill
    \includegraphics[width=0.30\textwidth]{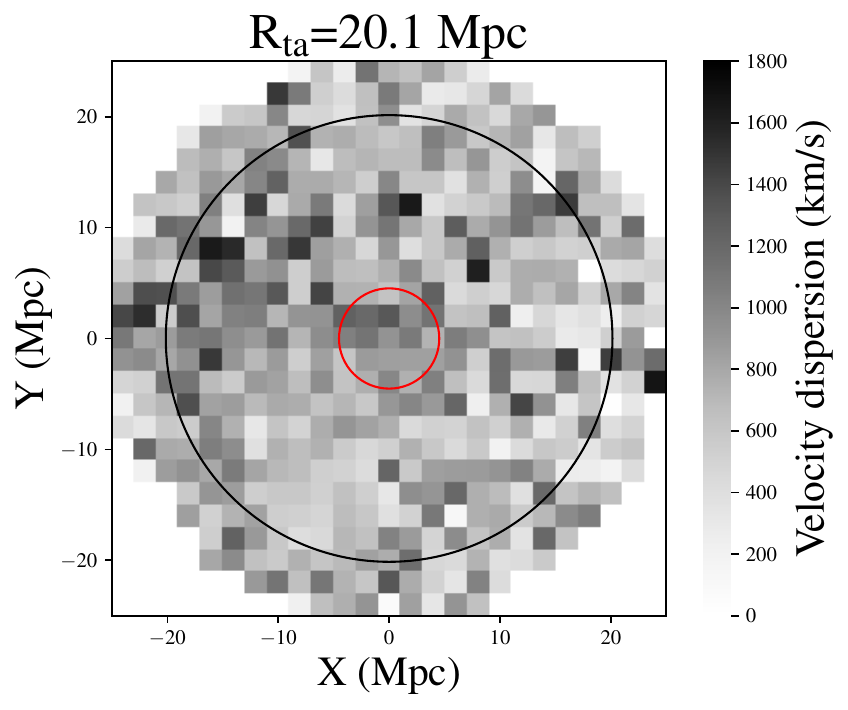}

    \caption{Stacked velocity dispersion images of the merged MDPL2 and Virgo $\rm \Lambda CDM$ data. The turnaround values of the dataset were segmented into nine bins. All of the corresponding clusters' projections of line-of-sight velocities, which belong in the same bin category, were stacked on top of each other to create nine images, calculating the standard deviation in each pixel. On top of each image is the mean value of the turnaround radii in each bin, which is represented with a black circle in each image. Red circles represent the mean $\rm R_{200}$ in each case. There is no apparent way of predicting the $\rm R_{ta}$ based on these images ``by eye,'' especially outside of the $\rm R_{ta}$ range $4.3-10.6 \; \rm Mpc$.}
    \label{fig: Stacked examples}
\end{figure*}

In Fig. \ref{fig: Stacked examples} we demonstrate that acquiring $\rm R_{ta}$ from stacked velocity profiles is not straightforward. When computing the velocity dispersion images based on velocity data from nine $\rm R_{ta}$ bins, the location of the turnaround scale is not identifiable ``by eye'' in the images.

\section{Stacked velocity dispersion based on $\rm R_{200}$}
\begin{figure*}[htb!]
    \centering
    \includegraphics[width=0.85\columnwidth]{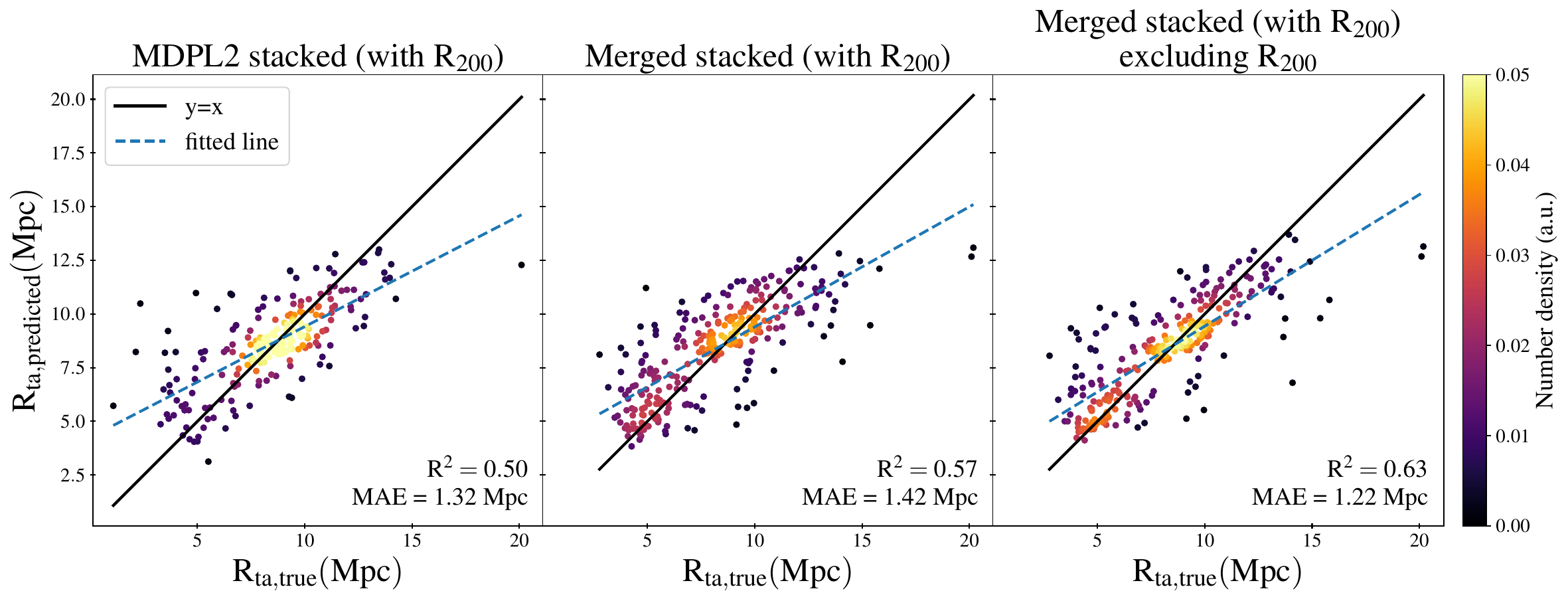}
    \hfill
    \caption{Comparison between predicted and true values of the $\rm R_{ta}$ for models using the stacked images from 3000 bins of $\rm R_{200}$ from: the MDPL2 data (left panel); the merged MDPL2 with Virgo $\rm \Lambda CDM$ data (middle panel); the merged MDPL2 with Virgo $\rm \Lambda CDM$ data with removed information inside the $\rm R_{200}$ of the central overdensity (right panel). $\rm R^2$ scores and the mean absolute errors (MAEs) are shown at the bottom right of each plot. The colors represent the number density of the plotted points, calculated using a Gaussian kernel density estimate. It is apparent that the merging of the two datasets significantly improves the performance of the model even without the central halos' velocity information.
    }
    \label{fig:velocity true predicted_r200}
\end{figure*}

Here (Fig. \ref{fig:velocity true predicted_r200}) we present the results that were computed in the same way as with Fig. \ref{fig:velocity true predicted}, but this time with stacking of the images based on the $\rm R_{200}$ of the clusters. The results of the middle and right panels seem to demonstrate similar success, while the left panel (only using MDPL2 data) seems to be better in this case. The reason behind this is probably due to better tracing of the global minimum during training.

\newpage 

\section{CNNs versus ``classical'' astrophysical approach}\label{sec:cnn vs classical}

We have established that significant information about the turnaround radius is encoded in the velocity dispersion of galaxies in our mock observations, in the form of the stacked images of our merged dataset (MDPL2 and Virgo $\Lambda$CDM). This was achieved by the somewhat unconventional use of 2D pixelized projections processed by CNNs. In the realm of astrophysics and cosmology, a more classical and commonly used approach would be to deduce information from the radial velocity profiles of the galaxy clusters. Here, we employed this method of more traditional compression, compared it with our previous results, and explored its limitations for generalizing to other redshifts.

\subsection{Baseline model}
As our more traditional baseline, we tried to train a simple model, based on the mass and velocity radial profiles of the projections, with the \texttt{XGBoost} algorithm \citep{chen2016xgboost}.
Here, instead of the $25 \times 25$ image creation described in Sect. \ref{sec:prepro} (2D array), we created a new dataset by calculating the mass column density and velocity dispersion in 25 radial bins in each projection (1D array). Consequently, each training instance had 25 features with the target variable being the turnaround radius.

\begin{figure}[htb!]
    \centering
    \includegraphics[width=0.33\columnwidth]{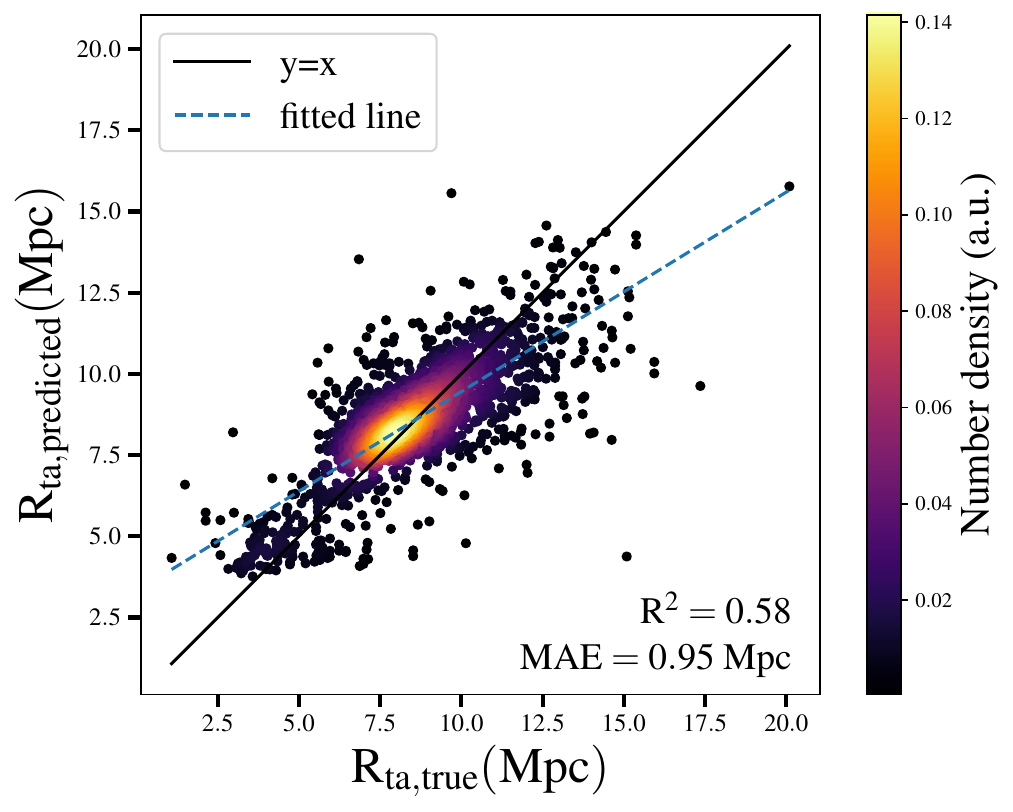}
    \includegraphics[width=0.33\columnwidth]{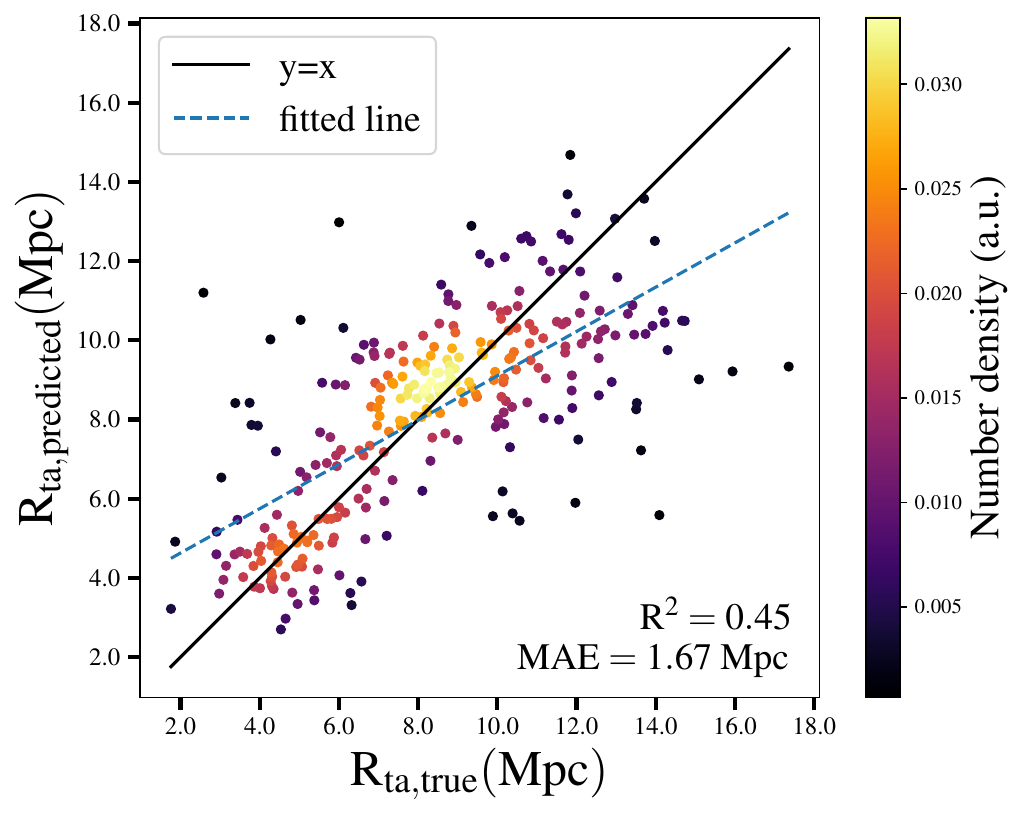} 
    \hfill
    \caption{Comparison between predicted and true values of the $\rm R_{ta}$ for 1D models (features from radial bins only) using MDPL2 mass images (left panel) and the merged MDPL2 and Virgo velocity dispersion images (right panel). The colors represent the number density of the plotted points, calculated using a Gaussian kernel density estimate.
    }
    \label{fig:mass radial true predicted}
\end{figure}

Using the mass column density MDPL2 data, the model performed with the same success, having an $\rm R^2$ score of $0.58$
as seen in the left panel of Fig. \ref{fig:mass radial true predicted}, where we plot the predicted versus true values in the same format as before. Again, this was an expected result since the model relied almost completely on the mass of the central cluster, which is not lost by the new type of compression.

A more interesting test is to compare the baseline model with the velocity dispersion data. 
After training it on the merged stacked dataset (which was previously the best candidate), testing resulted in an $\rm R^2$ score of $0.45$,
as seen in the right panel of Fig. \ref{fig:mass radial true predicted}.
In this case, the original CNN outperformed this model ($\rm R^2$ score: $0.57$, middle panel of Fig. \ref{fig:velocity true predicted}). 
The underlying reason for this behavior should be the more severe compression. This compression introduces a symmetry to our data that was not necessarily present in their original form. One important symmetry-breaking component is the inclusion of structures outside of the 25 Mpc radius from the velocity cut, which the CNN seems to be able to discard. 
It should be noted that a merged dataset with four copies of each projection, each rotated by 90 degrees, was performing worse, even with the use of the CNN, enhancing our previous argument.

\subsection{Redshift normalization}
\label{subsec:redshift_normalization}
At this point, the superiority of the original CNN is already substantiated from the previous section.
The only caveat was that our original model was not successful when tested on other cosmological settings. 
For this reason, we made a final attempt to counter this issue for the velocity dispersion data. 
In particular, we tried to normalize the velocity profiles in each redshift with the background Universe, in order to let the model generalize. 

Subsequently, we computed the mean value of the velocity dispersion of the last radial bin of all the projections of the training dataset in the same redshift.
Then, we tried two normalizations:
dividing by or subtracting the values of the bins from the mean value. 
We trained the model with the MDPL2 and the merged data on $z=0$ and tested on $z\approx0.5$ and $z\approx1.0$.
The resulting $\rm R^2$ scores can be seen in Table \ref{table 3}.

\begin{table}[htb!]
\centering
\caption{$\rm R^2$ scores of models trained on $z=0$ and tested on $z\approx0.5$ and $1.0$ for the MDPL2 and the merged data.}
\label{table 3}
\begin{tabular}{c c c }
\hline\hline
\noalign{\smallskip}

 & Divide by outer ring & Subtract outer ring \\ 
\hline
\noalign{\medskip}
MDPL2           \\ 
 \hline \noalign{\smallskip}
 
  $z=0$ & $0.17$ & $0.17$  \\
   $z\approx0.5$ &$-0.19$ & $-0.36$  \\
  $ z\approx1.0$ &$-7$ & $-8$  \\
   \hline \noalign{\medskip}
Merged         \\
\hline \noalign{\smallskip}

        $z=0$ & $0.45$ & $0.45$  \\
  $ z\approx0.5$ &$0.22$ & $0.17$  \\
  $ z\approx1.0$ &$-2.18$ & $-2.93$  \\
\hline
\end{tabular}
\end{table}

It can be seen that the model is not successful on other redshifts, even with these normalizations, and again exhibits the same behavior of the lower $\rm R^2$ score as we move away from the training redshift. However, the division of the outer ring in the merged dataset could suggest some potential for success, which requires further investigation beyond the scope of this work.

\end{appendix}

\end{document}